\documentclass[reprint,aps,prl, 
balancelastpage,
a4paper,floatfix,
superscriptaddress]{revtex4-2}

\usepackage[margin=0.75in]{geometry}
\usepackage[T1]{fontenc}
\usepackage{graphicx}
\usepackage{float}
\usepackage{amsmath,amssymb,bm}
\usepackage{xcolor}
\usepackage{booktabs}
\usepackage{enumitem}
\usepackage{dcolumn}
\usepackage{hyperref}
\usepackage{hycolor}
\hypersetup{colorlinks=true,citecolor=[rgb]{0,0.18,0.68}, 
urlcolor=[rgb]{0,0.18,0.68}, linkcolor=[rgb]{0,0.18,0.68}, final = true}
\usepackage[rightcaption]{sidecap}
\usepackage{physics}
\usepackage{mathtools}
\usepackage{relsize}
\usepackage{mathptmx}
\usepackage[scaled=0.86]{helvet}

\usepackage{siunitx}
\DeclareSIUnit\angstrom{\text {Å}}
         
  \def\be{\begin{equation*}}
  \def\ee{\end{equation*}}
  \def\ba{\begin{eqnarray}}
  \def\ea{\end{eqnarray}}

  \def\fref#1{Fig.~\ref{#1}}

  \def\bt{\textrm} 

  \def\nsi#1{\noindent\textit{\bt{#1.---}}}
  \definecolor{or}{RGB}{234,142,53}
  \definecolor{gr}{RGB}{150,150,150}
  \definecolor{bl}{RGB}{54,152,187}

  \def\rb{\mathbf{r}}
  \def\NN{\mathbf{D}}
  \def\dN{S}
  \def\dNi{S_{i}}

  \def\dNv{\mathbf{\dN}}
  \def\AF{\mathrm{AF}}
  \def\PDB{\mathrm{PDB}}
  \def\dNvP{\dNv^{\PDB}}
  \def\dNvA{\dNv^{\AF}}
  \def\AFave{\langle \AF \rangle}
  
  \def\RMSD{\mathrm{RMSD}}
  \def\corr{\mathrm{Corr}}
  \def\CA{\mathrm{C}_{\alpha}}
  
  \def\dm{\delta_\mathrm{m}}

  \newcommand{\eg}{\textit{e.g.}}

  \definecolor{YKB}{rgb}{0.00,0.18,0.65}

\begin{document}

\title{AlphaFold2 can predict single-mutation effects}

  \author{John M. McBride}
    \email{jmmcbride@protonmail.com}
    \affiliation{Center for Soft and Living Matter, Institute for Basic Science, Ulsan 44919, South Korea}
  \author{Konstantin Polev}
    \affiliation{Center for Soft and Living Matter, Institute for Basic Science, Ulsan 44919, South Korea}
    \affiliation{Department of Biomedical Engineering, Ulsan National Institute of Science and Technology, Ulsan 44919, South Korea}
  \author{Amirbek Abdirasulov}
    \affiliation{Department of Computer Science and Engineering, Ulsan National Institute of Science and Technology, Ulsan 44919, South Korea}
  \author{Vladimir Reinharz} 
    \affiliation{Universit\'{e} du Qu\'{e}bec \`{a} Montr\'{e}al, Canada}
  \author{Bartosz A. Grzybowski}
    \email{nanogrzybowski@gmail.com}
    \affiliation{Center for Soft and Living Matter, Institute for Basic Science, Ulsan 44919, South Korea}
    \affiliation{Departments of Physics and Chemistry, Ulsan National Institute of Science and Technology, Ulsan 44919, South Korea}
  \author{Tsvi Tlusty}
    \email{tsvitlusty@gmail.com}
    \affiliation{Center for Soft and Living Matter, Institute for Basic Science, Ulsan 44919, South Korea}
    \affiliation{Departments of Physics and Chemistry, Ulsan National Institute of Science and Technology, Ulsan 44919, South Korea}

\begin{abstract}
AlphaFold2 (AF) is a promising tool, but is it accurate enough to predict single mutation effects?  Here, we report that the localized structural deformation between protein pairs differing by only 1-3 mutations -- as measured by the effective strain -- is correlated across \num{3901} experimental and AF-predicted structures.  Furthermore, analysis of $\num{{\sim} 11000}$ proteins shows that the local structural change correlates with various phenotypic changes. These findings suggest that AF can predict the range and magnitude of single-mutation effects on average, and we propose a method to improve precision of AF predictions and to indicate when predictions are unreliable.
\end{abstract}

\maketitle

  Alteration of one or few amino acid residues can affect structure~\cite{tokco09,lorbi00,zhast18}
  and function~\cite{yanbi16,mcbmb22} of a protein and, in extreme cases, be the difference
  between health and disease~\cite{sahce15,prafm21}.
  Understanding structural consequences of point mutations is important for drug design~\cite{herpn13,albeo17}
  and could also accelerate optimization of enzymatic function via directed evolution~\cite{arnac98,caona22}.
  In these and other applications, theoretical models~\cite{stetb19}
  could be of immense help, provided they are sufficiently accurate.
  In this context, AlphaFold2~\cite{jumna21} has recently made breakthroughs in predicting
  global protein structure from sequence with unprecedented precision.
  Notwithstanding, it is not yet known whether AF is sensitive enough to
  detect small, local effects of single mutations.
  Even if AF achieves high accuracy, the effect of a mutation may be small
  compared to the inherent conformational dynamics of the protein -- predicting
  static structures may not be particularly informative~\cite{flejm21,morjm21,masjm21}.
  Furthermore, as accuracy improves, evaluating the quality of predictions becomes increasingly
  complicated by the inherent noise in experimental
  measurements~\cite{masjm21,andps07,rasac09,venci12,eveps16,hurps19,lynpa20,fowst21}.
  So far, no study has evaluated whether AF can accurately measure structural changes due to single mutations,
  and there are conflicting reports as to whether AF can predict the effect of a
  mutation on protein stability~\cite{pakpo23,buens22,akdns22,zhabi21,manbi22}.
  Furthermore, recent evidence suggests that AF learns the energy functional underlying folding, raising the question of whether the inferred functional is sensitive enough to discern the subtle physical changes due to a single mutation~\cite{ronpr22}.
  We aim to resolve this issue by comparing AF predictions with extensive data on protein structure and function.

  We examine AF predictions in light of structural data from a curated set of proteins from the Protein Data Bank (PDB)~\cite{berna00}, and phenotype data from high-throughput experiments~\cite{sarna16,poenc19,marsc21}. We find that AF can detect the effect of a mutation on structure by identifying local deformations between protein pairs differing by \numrange{1}{3} mutations. 
  The deformation is probed by the effective strain (ES) measure. We show that ES computed between a pair of PDB structures is correlated with the ES computed for the corresponding pair of structures predicted by AF.
  Furthermore, analysis of ${\sim}$\num{11000} proteins whose function was probed in three high-throughput studies shows significant correlations between AF-predicted ES and three categories of phenotype (fluorescence, folding, catalysis) across three experimental data sets~\cite{sarna16,poenc19,marsc21}.
  These sets of correlations suggest that AF can predict the range and magnitude of single-mutation effects.
  We provide new tools (\href{https://github.com/mirabdi/PDAnalysis}{github.com/mirabdi/PDAnalysis}) for computing deformation in proteins, and a methodology for increasing the precision of AlphaFold predictions of mutation effects.
  Altogether, these results indicate that AF can be used to predict physicochemical effects of missense mutations, undamming vast potential in the field of protein design and evolution.

 \begin{figure*}
\centering
\includegraphics[width=0.99\textwidth]{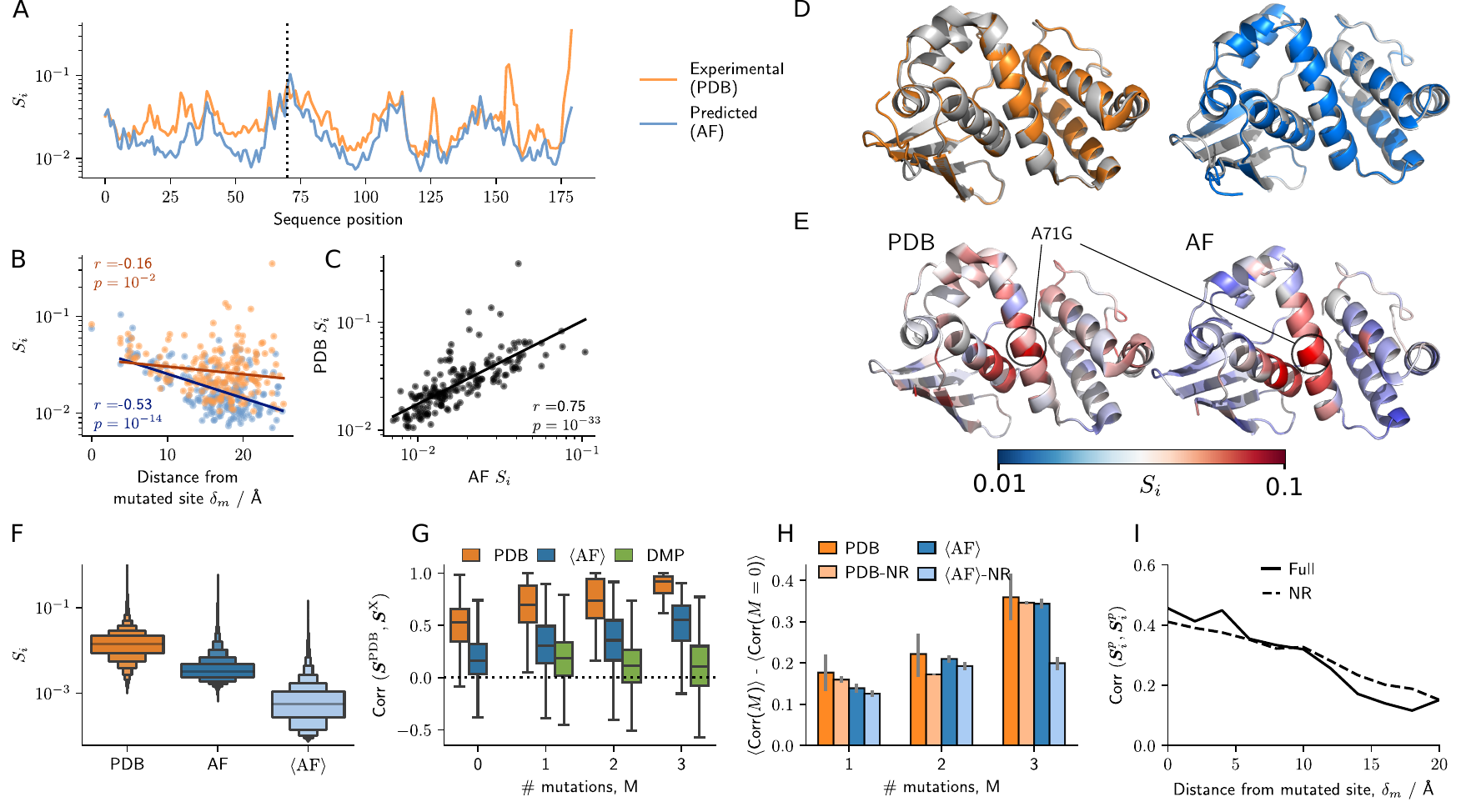}
\caption{\label{fig:fig1}
  A: Local deformation per residue measured by effective strain, $\dNi$, between wild-type (WT) and mutant (A71G) H-NOX protein, for experimental (orange) and AF-predicted (blue) structures. Dotted line indicates the mutated residue.
  B: $\dNi$ vs distance from the nearest mutated site, $\dm$.
  C: Comparison of $\dNi$ obtained from experimental and predicted structures.
  D: Overlaid WT (grey, \texttt{6BDD\_A}) and mutant (colour, \texttt{6BDE\_A}), experimental (orange) and predicted (blue) structures.
  E: Wild type protein with residues coloured by $\dNi$; location of A71G mutation is shown.
  F: Distribution of $\dNi$ between matched pairs of structures with the same sequence ($M=0$),
  for PDB, AF, and averaged AF ($\AFave$) structures.
  G: Distribution of correlation between PDB strain fields and equivalent fields from
  PDB, AF and DMPfold, shown for different numbers of mutations, $M$.
  H: Residual correlation that is due to mutations, shown for the full dataset and a non-redundant version (NR);
  whiskers show bootstrapped \SI{95}{\%} confidence intervals.
  I: Correlation between PDB and $\AFave$ strain fields, $\dNv^p_{i}$ , across all pairs $p$ and residues $i$ that are within a distance $\dm$ from a mutated site,
  shown for the full dataset and a non-redundant version (NR).
}
\end{figure*}

\nsi{AF can predict local structural change}
  We illustrate our approach by analyzing wild-type (WT; \texttt{6BDD\_A}) and single-mutant (\texttt{6BDE\_A}, A71G) structures of H-NOX protein from \textit{K. algicida} (\fref{fig:fig1}D)~\cite{hesac18}.
  To quantify local deformation, we calculate the effective strain (ES) per residue $\dNi$ (See App. A) for, respectively, experimental and AF-predicted pairs of structures (\fref{fig:fig1}A).
  The ES is the mean relative change in distance from $\CA$ of residue $i$ to neighboring $\CA$ positions within a range of \SI{13}{\AA}. ES provides a robust estimate of the magnitude of local strain, which accounts also for non-affine deformation in addition to affine deformation~\cite{mitpn16,eckprx17,dutpnas18,eckrmp19,falpr98}. 
  Like the frame-aligned-point-error (FAPE) measure used in training AF~\cite{jumna21}, ES is invariant to alignment.
  In H-NOX, we observe that the $\dNi$ is highest at, and decays away from the mutated site, showing a correlation with the distance from the mutated site (\fref{fig:fig1}B).
  We find that $\dNi$ is correlated across PDB and AF structures (\fref{fig:fig1}C,E).
  Taken together, these correlations suggest that $\dNi$ is a sensitive measure of local structural change, and that AF is capable of predicting such structural change upon mutation.

\nsi{Experimental measurement variability limits evaluation}
  Before exploring AF predictions in more detail, we first examine variation within experimental structures by comparing repeat measurements of the same protein.
  In \fref{fig:fig1}F we show the distribution of $\dNi$ calculated for all residues in all
  pairs (Supplemental Material (SM) Sec. 1A~\cite{supp}) of protein structures with identical sequences (number of mutations, $M=0$);
  we excluded pairs where the crystallographic group differed (SM Sec. 1B~\cite{supp}).
  Protein structures vary considerably between repeat measurements (average ES is $\langle\dNi\rangle = 0.018$, and the average Root Mean Square Deviation is $\RMSD = \SI{0.24}{\angstrom}$).
  In comparison,
  differences between repeat predictions of AF are much lower ($\Delta\dNi = 0.005$, $\RMSD = $\SI{0.11}{\angstrom}). 
  For example, the experimental RMSD between WT and mutant H-NOX 
  is \SI{1.6}{\AA}, while the AF-predicted RMSD is \SI{0.3}{\AA}.
  We can refine AF predictions further by making multiple repeat predictions and averaging
  over the local neighborhoods ($\AFave$ in Fig.~\ref{fig:fig1}F, App. B), which results in even lower differences ($\Delta\dNi =0.001$). We find that averaging decreases deformation away from mutated residues, while preserving deformation in mutated areas (SM Sec. 6~\cite{supp}), thus we henceforth report results for averaged structures, except where noted.
The variation between experimental measurements might mask the deformation due to mutation, and therefore limits our ability to evaluate AF predictions.

\nsi{Mutation effects are measurable in PDB structures}
  To quantify how well we can measure mutation effects from PDB structures, we compare
  deformation between two matching pairs of PDB structures with identical ($M=0$)
  and non-identical ($M>0$) sequences (SM Sec. 5B~\cite{supp}) of length $L$ (number of residues). For each pair, we calculate the strain fields, $\dNv = (\dN_1,\ldots, \dN_L)$, which record ES values for all residues,
  and we calculate Pearson's correlation coefficient $r$ as in \fref{fig:fig1}C.
  We find that even among protein structures with identical sequences, strain fields are highly correlated (\fref{fig:fig1}G). This occurs because
  the magnitude of positional fluctuations depends on local flexibility;
  more flexible regions exhibit higher strain in repeat measurements (App. B).
  Thus, a portion of the $\dNv$ correlation in \fref{fig:fig1}C is due to effects
  other than mutation. Despite this, we find that correlations are much higher when comparing
  pairs of structures that differ by one or more mutations ($M>0$), and correlations increase 
  with $M$ (\fref{fig:fig1}G). Thus, the strength of PDB-PDB deformation correlations
  is partly due to differences in local flexibility, and partly also due mutations.

\begin{figure*}
\centering
\includegraphics[width=0.99\textwidth]{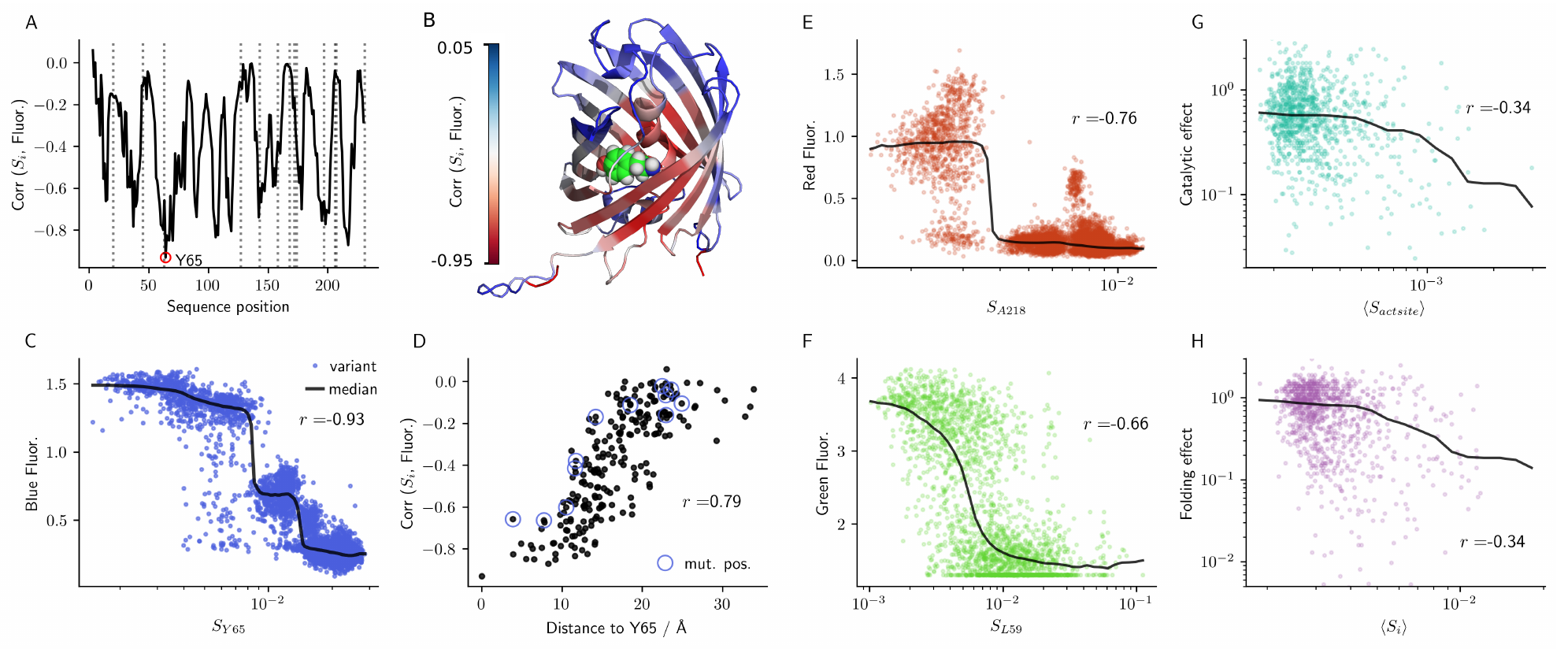}
\caption{\label{fig:fig2}
  A: Correlation (Pearson's $r$) between blue fluorescence (mTagBFP2) and AF-predicted effective strain (ES), $\dNi$, between WT and \num{8191} variants for all sequence positions $i$; positions of mutated residues are shown by dotted lines;
  chromophore site (Y65) is indicated (red circle).
  B: Structure of BFP, with each residue colored according to $\corr(\dNi, \textrm{Fluor.})$ (A); Y65 atoms are shown as spheres.
  C: Strain at residue Y65 vs. fluorescence for mTagBFP2 variants.
  D: Fluorescence-strain correlation per residue vs. distance from residue $i$ to Y65; mutated positions are indicated (blue circle).
  E-H: Correlations between: $\dN_{A218}$ and red fluorescence (mKate2); $\dN_{L59}$ and green fluorescence (GFP); catalytic activity and $\dN$ at the active site (PafA); folding ability (fraction of active enzymes) and average strain, $\langle \dNi \rangle$, of the \num{50} residues that correlate best with folding ability (PafA).
 }
\end{figure*}

\nsi{Mutation effects are correlated across PDB and AF structures}
  To evaluate the performance of AF in predicting mutation effects, we calculate correlations
  between PDB and AF-predicted strain fields, $\dNvP$ and $\dNvA$, calculated for
  all matched pairs of proteins (SM Sec. 5B~\cite{supp}). The PDB-$\AFave$ correlations between pairs of structures with identical sequences ($M=0$)
  are lower than PDB-PDB correlations (\fref{fig:fig1}G), as are the correlations for
  non-identical sequences ($M>0$). 
  Nonetheless, the correlations are significant
  and they increase with $M$. To put this result in context, the PDB-AF correlations are
  considerably higher than correlations obtained by using another
  algorithm to predict protein structure (DMPfold2)~\cite{kanpn22}.
  To compare the degree of correlation that is due to mutation effects,
  we plot the mean correlation for non-identical sequences
  $\langle \corr\big(M\in\{1,2,3\}\big) \rangle$ subtracted from the mean correlation
  that can be attributed to fluctuations, $\langle \corr\big(M = 0\big) \rangle$.
  \fref{fig:fig1}H shows that the degree of correlation due to mutations is as high for
  AF-PDB comparisons as it is for PDB-PDB comparisons.
  Since many protein families are over-represented in the PDB, we repeat the analyses
  on non-redundant sets of proteins (SM Sec. 1C~\cite{supp}), finding that AF-PDB correlations are still
  comparable to PDB-PDB correlations (NR in \fref{fig:fig1}H).

\nsi{AF predicts the range of mutation effects}
  \fref{fig:fig1}G-H shows that within matched protein pairs, deformation is correlated between PDB and AF, although the magnitude of deformation can differ (\fref{fig:fig1}F). This
  indicates that AF is at least correctly predicting the range and the relative strength of the effect of a mutation. On average, AF predicts that mutations can produce changes in structure up to \SIrange{16}{18}{\AA} (SM Sec. 7~\cite{supp}), whereas the average range in the PDB data is only\SI{14}{\AA} due to the higher measurement variance in the PDB. This suggests that AF correctly predicts the range of a mutation's effect on structure.

\nsi{AF predicts the relative magnitude of mutation effects}
  It is essential to be able to predict whether a mutation will lead to a big or small effect on structure. While the previous analysis did not show this, we directly address this problem by examining whether predicted effects correlate with empirical
  effects \textit{across} proteins. To do this, we group $\dNi$ values from all matched pairs $p$ by
  distance from the nearest mutated residue, $\dm$ (in bins of \SI{2}{\AA}), to get sets of $\dNv^p_i$
  for both PDB and $\AFave$ pairs of structures. This allows us to compare 
  ES magnitudes across proteins, by calculating the correlation between $\dNv^p_i$ for PDB and $\AFave$. At mutated sites, the correlation is quite high, and decreases away from the mutated site as expected (\fref{fig:fig1}I);
  this is also true for the non-redundant sample. Hence, AF is capable of distinguishing between mutations that have relatively large or small effects on structure.

\nsi{Phenotypic change correlates with AF-predicted ES}
  An orthogonal test of whether AF can predict the effect of a mutation is to study correlations between the effective strain (ES), $\dNi$, and phenotypic change. This approach avoids the pitfalls associated with noisy PDB measurements, and allows us to test predictions of structures that AF was not directly trained on.
  However, the link between structure and function is often unknown. The mapping from genotype to phenotype is complex and involves dimensional reduction \cite{mcbmb22,tlupt16,eckbi21}. Therefore, a lack of a correlation between  $\dNi$ and phenotype is not strong evidence that the structure is incorrect, as there maybe be a non-trivial mapping between structure and function. 
  On the other hand, observation of correlations between $\dNi$ and phenotype is strong evidence that AF can be predictive in estimating the effect of mutations.
  We study three data sets from high-throughput experiments, covering three
  distinct phenotypes (SM Sec. 2~\cite{supp}):
  (i) blue and red fluorescence is measured for \num{8192} sequences linking mTagBFP2
  (blue) and mKate2 (red)~\cite{poenc19};
  (ii) green fluorescence is measured for \num{2312} GFP sequences~\cite{sarna16};
  (iii) folding (fraction of active enzymes) and catalytic ($k_\mathrm{cat}$) effects of mutations are measured for PafA~\cite{marsc21} (SM Sec. 2C~\cite{supp}).

  We find significant correlations between phenotype and AF-predicted ES (compared to WT) for all phenotypes (\fref{fig:fig2}).
  It is possible to predict blue, red and green fluorescence (Pearson's $r=-0.93$, $r=-0.76$, $r=-0.67$) by measuring the ES at residues Y65, A218 and L59, respectively (\fref{fig:fig2}C,E,F).
  There are many other residues at which deformation measured by ES is predictive of fluorescence (\fref{fig:fig2}A-B), and these residues are found to be closer to residue Y65 (\fref{fig:fig2}D, Y65 covalently binds to a chromophore); this is despite no mutations to Y65, which suggests that AF can predict allosteric effects. We also find weaker, yet significant correlations between ES and the empirical effects of mutations on folding and catalytic activity (\fref{fig:fig2}G-H). For catalytic activity,   we measure mean deformation at the active site; for the folding effect, we measure mean ES between the 50 residues that correlate best with the folding effect (SM Sec. 9~\cite{supp}).
  
  In contrast, we do not find consistent correlations with RMSD, a standard estimate of AF accuracy~\cite{jumna21}, indicating that local deformation, as measured by the ES, is more appropriate for measuring mutational effects (SM Sec. 10). In some cases, performance is heavily dependent on which pre-trained model (SM Fig. 4~\cite{supp}) is used: surprisingly, we found that using the highest ranked (by pLDDT; see SM Sec. 3 \& 11~\cite{supp}) models resulted in worse performance for phenotypic change (SM Fig. 4~\cite{supp}), and performance for structural change was close to average (SM Sec. 12~\cite{supp}). Taken together, these results provide evidence that AF can be used to predict the structural effect of a single mutation.

\nsi{ES correlates with phenotypic change for wild-type proteins}
  It is quite unexpected that ES, $\dNi$, should be a good predictor of phenotypic change, even if AF can accurately predict structure. We suspect that the correlation is strong because the structures are always compared to the wild-type (WT) proteins, where the structure is adapted for function through evolution -- any deviation from this optimal structure is likely to diminish protein function. We find that high correlations are only found within $M \leq 8$ mutations from the WT, and phenotype-ES correlations are much weaker between non-WT pairs (SM Sec. 9~\cite{supp}). Thus we conclude that $\dN_i$ is a good predictor of phenotypic change from native protein sequences. For studying phenotypic change away from optima in phenotype landscapes, another mapping from structure to function is needed.

\nsi{Discussion}
  We have shown that AF is capable of predicting structures with sufficient accuracy and that it can pick up changes as small as those resulting from a single missense mutation. Direct validation
  of predicted mutational effects on structure is limited by the accuracy of empirical structures (\fref{fig:fig1}F), and further hindered by the lack of sequence pairs that are suitable for comparison (SM Sec. 1~\cite{supp}).
  Likewise, predicting phenotypic change from structure alone ought to be challenging, to say the least.
  Despite these steep hurdles, we have shown, using effective strain (ES) as a measure of deformation,
  that differences between AF-predicted structures do correlate with both structural (\fref{fig:fig1}) and phenotypic changes (\fref{fig:fig2}) in empirical data.
  Examining individual pairs of PDB structures, mutation effects are masked by fluctuations, but this inherent noise is filtered by analyzing the statistics of many pairs, demonstrating that AF is accurate.
  The difficulties in assembling sufficient data for validation highlight that the age of
  experimental protein structure identification is far from over~\cite{terbi22}, despite the success of
  AF and RoseTTAFold~\cite{jumna21,baesc21}. Our methodology for evaluating mutation effects using deformation can be used in future empirical evaluation of mutation effects.

\nsi{Advice for using AF to study mutations}
  We find higher correlations between AF and PDB when mutations are in less flexible
  regions of proteins, and when mutations have large effects (App. C). One can quickly estimate
  flexibility using pLDDT (AF's confidence in a residue's predicted position,
  or a proxy measure of rigidity; SM Sec. 11~\cite{supp}), but it is more useful to measure the variance of AF predictions
  by predicting multiple structures (App. B, SM Sec. 6~\cite{supp}). Depending on the flexibility, and
  mutation effect size, one can achieve much more reliable estimates of mutation effects
  by averaging across many repeat structures.
  We advise against using templates in predictions (used by default in AF models 1 and 2),
  since this appears to offer at best negligible increases in accuracy, and we found
  one example where including templates made the predictions much worse (SM Sec. 12~\cite{supp}).
  We recommend using effective strain as a measure of local deformation, rather than using RMSD or pLDDT. We provide code for calculating deformation, producing average structures, and calculating repeat-prediction variance at \href{https://github.com/mirabdi/PDAnalysis}{github.com/mirabdi/PDAnalysis}.

\nsi{AF predicts structure, not folding}
  We need to emphasise that AF is only trained to predict structures of stable proteins, and we make no claims about whether the proteins will indeed fold into the predicted structure. Given the marginal stability of most proteins, mutations may easily destabilize a protein so that its melting temperature falls below room temperature. The process of protein folding is carefully tuned \textit{in vivo} for folding on the ribosome, and through interactions with chaperones, and mutations that do not change structure may retard folding through other mechanisms \cite{mcbbj2021}. To see whether pLDDT is predictive of whether a protein will fold or not, we studied a set of 147 WW-domain-like sequences, of which 40 were found to fold \textit{in vitro}.
  Although more sophisticated methods may perform better, mean pLDDT by itself proved insufficient to sort folding from non-folding proteins (SM Sec. 11B~\cite{supp}).
  Now that one question -- what structure will a protein likely fold into? -- has been seemingly solved, at least partially, it is crucial to next answer the question of \textit{whether} a protein will spontaneously fold.

\nsi{Local deformation should be used to measure mutation effects}
  Placing the current results in a broader context, we note that the evidence in support of AF's capacity to predict the effect of a mutation has so far been mixed.
  Some studies suggest that AF and  RoseTTTaFold can be indirectly used to predict phenotype, but not by comparing structures~\cite{akdns22,zhabi21,manbi22}.
  Two studies have reported negative results~\cite{pakpo23,buens22}, which we attribute primarily to their use of pLDDT and RMSD -- measures much less precise of mutational effects compared to strain (SM Sec.~10-11~\cite{supp}). In one study, the authors found only weak correlations between pLDDT and fluorescence using the same GFP dataset used here.
  Although we do not expect pLDDT to strongly correlate with fluorescence, we do find higher correlations than those reported in~\cite{pakpo23} by examining allosteric effects (SM Sec.~11A~\cite{supp}).
  In another analysis~\cite{buens22}, the authors appear to assume that structure-disrupting mutations should result in a large change in predicted structure or pLDDT~\cite{buens22}.
  We first note that this paper only studied three proteins, limiting our ability to draw general conclusions. We also see that the deformation due to mutations in one of these proteins is higher than \SI{96}{\%} of mutation effects in our PDB sample (SM Sec. 13~\cite{supp}); 
  it is possible that such large deformation is predictive of destabilization, and testing this is a promising future direction \cite{McBride2023}.
  Ultimately, we think the present study has demonstrated that deformation (measured by ES) is a more robust measure of structural change upon mutation.

\nsi{Limitations}
  Our structural analysis is limited to showing statistical correlations, and more precise experimental measurements are needed to validate the prediction accuracy of single proteins. Likewise, we are limited to evaluating structural change in the actual training data, but a less biased evaluation may become possible as more mutation effects are empirically determined. Further work is needed to more extensively examine the effects of MSA coverage and depth on mutation prediction accuracy. As for the phenotypic effect, we analyzed two protein folds and three phenotypes, a this analysis ought to be replicated on a greater variety of proteins and phenotypes.

  In summary, we showed here that AF predictions of local structural change, probed by strain \cite{mitpn16,eckprx17,dutpnas18,eckrmp19}, can be used to study missense mutations in proteins. These analyses suggest that AF can, indeed, be a powerful tool, if used in the right context and backed up by appropriate analyses. Using AF, we can bridge the gap between sequence and function in high-throughput deep-mutational scan experiments, guide directed evolution studies~\cite{arnac98}, and design drugs \textit{in silico}~\cite{caona22}. For example, on a smaller scale, AF can be used to screen potential mutants, and in costly experiments where the number of mutations is limited, one can select mutations with strong or weak effects in desired regions of the protein.
  Overall, it appears that AF provides a step change in our ability to study and guide protein evolution. 
  All AF structures analyzed here are available at \cite{zenodo}. PDB files were compressed using Foldcomp \cite{foldcomp}.\\

  We thank Jacques Rougemont, Jean-Pierre Eckmann, Martin Steinegger and Milot Mirdita for discussions. We thank Jacque Rougemont for providing code to calculate shear.
  This work was supported by the Institute for Basic Science (IBS-R020-D1).\\

\nsi{Appendix A: Calculating local deformation}
  As a measure of local deformation, we compute the effective strain (ES), $\dNi$. ES is simply the mean relative change of the inter-particle distances around a given residue and is partially correlated with shear strain (SM Sec. 4-5~\cite{supp}). To calculate $\dNi$ per residue $i$, we first define a neighborhood $N_i$ that includes the $n_i = |N_i|$ residues $j \in N_i$ whose $\CA$
  positions $\rb_j$ are within \SI{13}{\AA} of $\rb_i$, the $\CA$ position of residue $i$ (in both reference and target structures).
  We obtain a $3\times n_i$ \textit{neighborhood} tensor $\NN_i$ whose $n_i$ rows are the distance vectors,
  $\rb_{ij} = \rb_j - \rb_i$.
  We calculate, respectively, $\NN_i$ and $\NN_{i}'$ for the two structures we are comparing
  (\eg, WT and mutant), and rotate $\NN_{i}'$ to maximize overlap between the tensors.
  The ES is the average over the $n_i$ neighbors of the relative change in the distance vectors,
  \begin{equation}
  \dNi = \left\langle \frac{|\Delta \rb_{ij}|}{|\rb_{ij}|} \right\rangle = \frac{1}{n_i}\sum_{j \in N_i} \frac{| \rb_{ij} - \rb_{ij}' |}{| \rb_{ij}|}~.
  \end{equation}
  We have evaluated several other local metrics, similar in nature to ES, finding that the conclusions are not very sensitive to the specific choice of metric
  or neighborhood cutoff (SM Sec. 4-5~\cite{supp}).
  We only include AF-predicted residues in strain calculations if pLDDT $> 70$, and treat them as disordered otherwise.

\begin{figure}
  \centering
  \includegraphics[width=0.45\textwidth]{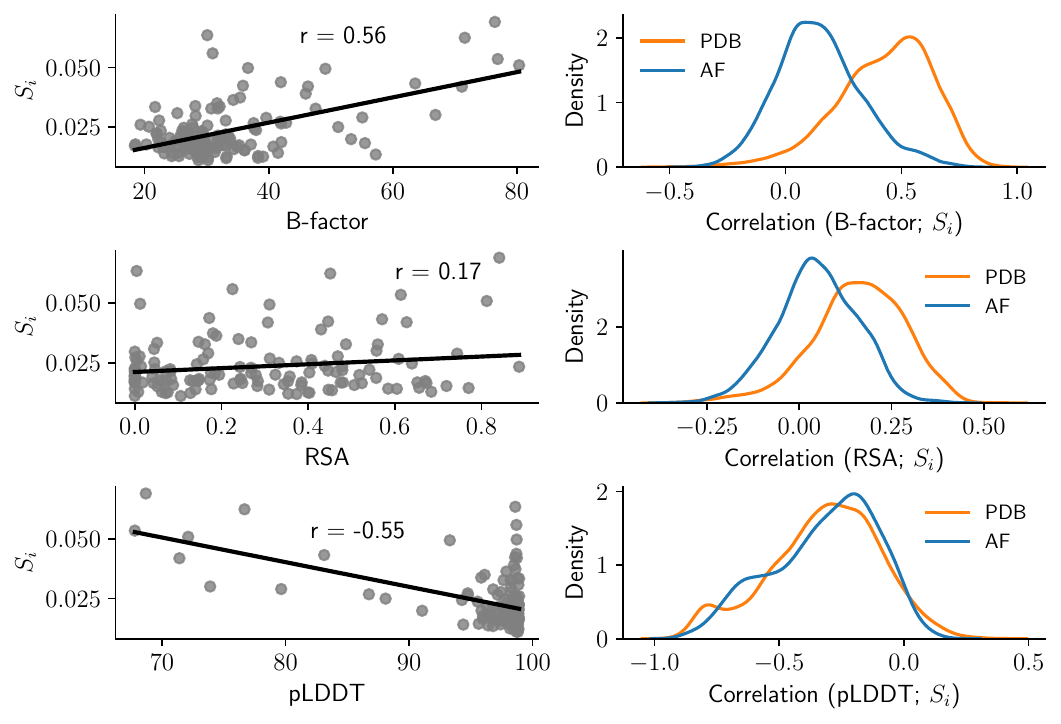}
  \caption{\label{fig:app1}
  \textbf{Fluctuations are greater in flexible regions.}
  Deformation (ES) between experimental hen lysozyme structures (\texttt{194L\_A} and \texttt{6RTA\_A}),
  $\dNi$ is correlated (Pearson's $r$) with B-factor, relative solvent accessibility (RSA), and pLDDT (left). Distributions (kernel density estimates)   of correlations for all proteins (right).
}
  \end{figure}

\nsi{Appendix B: Averaging local neighborhoods increases accuracy}
  Since AF predictions are stochastic, repeat predictions vary. We find that deformation between repeat predictions of the same protein leads to non-negligible ES (\fref{fig:fig1}G). The ES is higher in flexible regions, which is indicated by higher B-factor, solvent accessibility (RSA), and lower pLDDT (\fref{fig:app1}).
  It is possible to obtain more reliable estimates of mutation effects by averaging across local neighborhoods, $\NN_i$, in repeat predictions (SM Sec. 6~\cite{supp}). 
  Our average structures ($\AFave$) are typically averaged over all 5 AF models, with one set of predictions from DeepMind's AF implementation, and five sets of predictions from ColabFold's AF implementation~\cite{mirnm22}.
  Averaging typically increases deformation-phenotype correlations (SM Sec. 6B~\cite{supp}). One exception is the  mTagBFP2/mKate2 dataset, where DeepMind's implementation of AF produces a better correlation than the average; we find that this is due to the ColabFold implementation performing poorly on this specific protein (SM Sec. 6B~\cite{supp}).
  We see little increase in PDB-AF structure correlations (SM Sec. 6A~\cite{supp}), and we attribute this to limitations of higher repeat-measurement variability in PDB structures.

\begin{figure}
  \centering
  \includegraphics[width=0.45\textwidth]{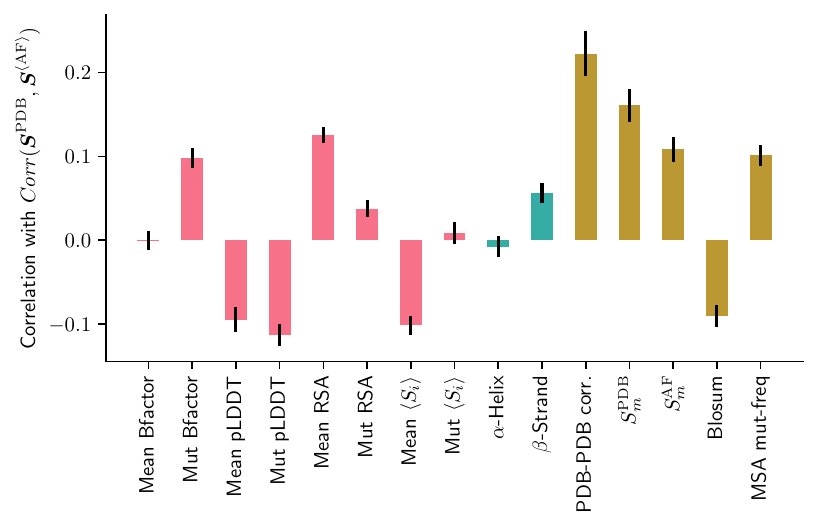}
  \caption{\label{fig:app2}
  \textbf{Correlations are higher if mutations have large effects in rigid regions.}
  A: Pearson's correlation between PDB-AF $\dNv$-correlation and:
  mean and mutated residue values of flexibility (B-factor, RSA, pLDDT, $\langle \dNi \rangle$);
  fraction of secondary structure ($\alpha$-helix or $\beta$-sheet); magnitude of mutation effect (PDB-PDB $\dNv$ correlation, ES at mutated site in PDB and AF, $S_\mathrm{m}^\mathrm{PDB}$ and $S_\mathrm{m}^\AF$, BLOSUM score, frequency of mutation in MSA). Results are shown for the non-redundant sample; whiskers show bootstrapped standard deviations.
  }
  \end{figure}

\noindent\textit{Appendix C: When do AF predictions correlate with PDB data?}~---
  Here we assess why AF sometimes predicts mutation effects similar to those measured in experimental structures (\fref{fig:fig1}G).
  Across all proteins, AF-PDB correlations are higher for mutant pairs of proteins in two situations (\fref{fig:app2}, SM Sec. 8~\cite{supp}): 
  when flexibility is low (low B-factor, low RSA, high pLDDT, high ES when comparing repeat predictions $\langle \dNi\rangle$);
  and when mutations have large effects that are easier to measure (high PDB-PDB correlation, high deformation at mutated site $S_\mathrm{m}$, BLOSUM score). One might expect a negative correlation with the frequency of mutation in MSA, as more frequent mutations might have smaller effects; instead, it appears that wider MSA coverage leads to more evolutionary information that improves predictions, but this needs to be tested further. There was no significant effect due to secondary structure type or MSA size (SM Fig. 21~\cite{supp}).

\nocite{libi06,kuf12,kabac76,zhapr04,zemna03,marbi13,lub08,suncr19,guosr22,maps23,zhaps09,delel22,salbi22,kabbi83,henpn92,piops22,socna05,matscipy}
\bibliography{af2_eval}

\end{document}


\title{
{\smaller[0.5] Supplemental Material for} \\ ``AlphaFold2 can predict single-mutation effects''}

  \author{John M. McBride}
    \email{jmmcbride@protonmail.com}
    \affiliation{Center for Soft and Living Matter, Institute for Basic Science, Ulsan 44919, South Korea}
  \author{Konstantin Polev}
    \affiliation{Center for Soft and Living Matter, Institute for Basic Science, Ulsan 44919, South Korea}
    \affiliation{Department of Biomedical Engineering, Ulsan National Institute of Science and Technology, Ulsan 44919, South Korea}
  \author{Amirbek Abdirasulov}
    \affiliation{Department of Computer Science and Engineering, Ulsan National Institute of Science and Technology, Ulsan 44919, South Korea}
  \author{Vladimir Reinharz} 
    \affiliation{Universit\'{e} du Qu\'{e}bec \`{a} Montr\'{e}al, Canada}
  \author{Bartosz A. Grzybowski}
    \email{nanogrzybowski@gmail.com}
    \affiliation{Center for Soft and Living Matter, Institute for Basic Science, Ulsan 44919, South Korea}
    \affiliation{Departments of Physics and Chemistry, Ulsan National Institute of Science and Technology, Ulsan 44919, South Korea}
  \author{Tsvi Tlusty}
    \email{tsvitlusty@gmail.com}
    \affiliation{Center for Soft and Living Matter, Institute for Basic Science, Ulsan 44919, South Korea}
    \affiliation{Departments of Physics and Chemistry, Ulsan National Institute of Science and Technology, Ulsan 44919, South Korea}

\maketitle
\tableofcontents

  \section{PDB Structure Data}
    \subsection{Full}
  We curate a set of structures from the PDB (downloaded 10 August 2020) to study the effect of mutations on protein structure.
 We first select all proteins from the PDB that have equal sequence length, for which there are multiple structures whose sequences differ by no more than \num{3} mutations; we include proteins with identical sequences as a control group.
  We only include proteins of length $50 \leq L \leq 500$.
  Binding can result in large structural change, so we control for this:
  we exclude all protein complexes, or proteins bound to RNA or DNA since large interactions lead to large deformations with relatively poor reproducibility across repeat measurements;
  we only match pairs of proteins if they are bound to the exact same types of small molecules, since these result in small, more reproducible deformations.
  We exclude NMR structures for simplicity, avoiding the need to determine additional cutoffs to infer disorder, or to choose a representative structural model; we also found that NMR structures were much less reliable (higher deformation between repeat measurements) than crystal structures.
  We only consider pairs that were prepared at a similar pH (within $\pm 0.5$).
  We exclude pairs of structures that are almost identical (\eg, from time-resolved crystallography experiments); \ie, pairs with a RMSD smaller than \SI{0.001}{\angstrom}.
  This leaves us with \num{3901} PDB structures, and $\sim$\num{90000} pairs; $\sim$\num{70000} of these pairs derive from a set of \num{485} endothiapepsin structures, so we remove most of these until we are left with \num{17813} pairs.

  \subsection{Differences in crystallographic group}
  In about \SI{10}{\%} of pairs, the crystallographic groups are different.
  We find that this does not affect the PDB-AF deformation correlations.
  We do find that it affects the absolute values of deformation, so
  we do not include these pairs in the distribution of $\dNi$
  in the main text Fig. 1F.

  \subsection{Non-redundant sets}
  To create a non-redundant sample, we cluster protein sequences using CD-hit~\cite{libi06}, with a \SI{90}{\percent} sequence identity threshold.
  From the total pool of pairs, we create sub-samples with no more than \num{10} examples per group, for each value of $M$, the mutation number (in total, \num{2636} pairs); to estimate sampling error, we re-run analyses with \num{1000} sub-samples.

  \subsection{Structures not used in training AF}
  A major limitation of evaluating PDB-AF correlations is that AF was trained on the PDB structures that we use in evaluation. In our dataset, we found only \num{211} out of \num{17813} cases where all structures in a matched set were deposited in the PDB after the cutoff used for the training set (28 August 2019). 
  Out of these, \num{153} are structures of bovine trypsin. Hence, it is not currently possible to evaluate AF's performance on structures that it was not trained on, due to insufficient examples in this dataset.

  \section{High-throughput Phenotype Data}\label{sec:phendata}
  \subsection{eqFP611}
  There are two variants of the fluorescent protein epFP611, mKate2 and mTagBFP2, which exhibit respectively red and blue fluorescence. These two proteins differ by only \num{13} mutations. 
  A previous study measured blue and red fluorescence
  for all $2^{13}$ sequences that account for both mKate2, mTagBFP2, and all
  intermediate sequences~\cite{poenc19}. When comparing deformation with phenotype, we assign
  mKate2 to be the wild-type (WT) for red fluorescence, and mTagBFP2 to be
  the WT for blue fluorescence. We then calculate deformation of each variant
  compared to the WT, for each fluorescence colour. 
  In the main text, we report deformation calculated using single structures using DeepMind's implementation of AF (\ref{sec:ave}).
  
  \subsection{GFP}
  The GFP dataset is a subset (\num{2312} sequences) of the full dataset
  (\num{51715} sequences) published in \cite{sarna16}. The WT sequence
  was used to create a library of variants via random mutagenesis, and green fluorescence
  was measured for WT and all variants. The highest number of missense mutations is $M = 15$,
  although the average is much lower (\num{3.7}). To create a subset of the
  full dataset, we first grouped sequences by the number of mutations $M$,
  and picked at most \num{200} variants per value of $M$. We next chose variants
  in order to maximise the variance in fluorescence: We marked fluorescence values
  on a \num{200}-point, equally-spaced grid from the minimum to the maximum fluorescence (for each value of $M$), and picked the variants
  with fluorescence closest to each grid point. Due to the sparsity of fluorescence
  values between \num{2} and \num{2.5}, many grid points were assigned the same
  variant. We only included each variant once, and randomly chose variants to
  make up the remainder if there were less than \num{200} unique variants.
  When comparing deformation with phenotype, we calculate deformation
  of each variant compared to the WT.
  In the main text we report deformation calculated using averaged AF structures (\ref{sec:ave}).

  \subsection{PafA}
  The PafA dataset contains the WT phosphate-irrepressible alkaline phosphatase (PafA) of Flavobacterium, and 1036 mutants~\cite{marsc21}. All possible single-mutants involving substitutions to glycine or valine were created; if the native amino acid was glycine or valine, then the substitution was to alanine; one double mutant at the active site was also produced.
  For each protein, the catalytic rate constant $k_\mathrm{cat}$, Michaelis constant $K_\mathrm{m}$, and catalytic efficiency $k_\mathrm{cat}/K_\mathrm{m}$ were measured, alongside expression levels (using a GFP tag).
  To help infer whether mutations affected protein folding, the authors measured the kinetic constants at different temperatures and Zn concentrations, for different substrates, and with inhibitors. 
  They then separated the kinetic measurements into components of both catalytic effects and folding effects (which they term the fraction of active enzymes). When comparing deformation with phenotype, we calculate deformation of each variant compared to the WT.
  In the main text, we report deformation calculated using averaged AF structures (\ref{sec:ave}).

  \section{Structure Prediction}
  \subsection{AlphaFold: DeepMind}
  We predict structures using AF with a default template cutoff date (14 May 2020) and a reduced genomic database.
  We run AF using all five pre-trained models. Models \num{1} and \num{2} use structural templates as input;
  see \sref{sec:model} for more detail on differences between models.

  \subsection{AlphaFold: ColabFold}
  After initially using the DeepMind's implementation of AlphaFold, we switched to using the ColabFold implementation
  as it is faster and allows more control over the internal parameters of the algorithm~\cite{mirnm22}.
  We used ColabFold to produce five repeat predictions of structures for each sequence, for all
  five AF models. This resulted in \num{25} ColabFold structures per sequence in addition to
  the \num{5} AlphaFold structures per sequence. We ran ColabFold without templates, and
  used \num{6} recylces per structure.

  \subsection{DMPfold}
  To put the results of AF in context, we also predict structures using DMPfold~\cite{kanpn22}.
  We chose DMPfold primarily because it is fast. Note that DMPfold is a deterministic algorithm
  so there is no repeat-prediction variability.

  \begin{figure*}[t!]  \centering
  \includegraphics[width=0.99\linewidth,]{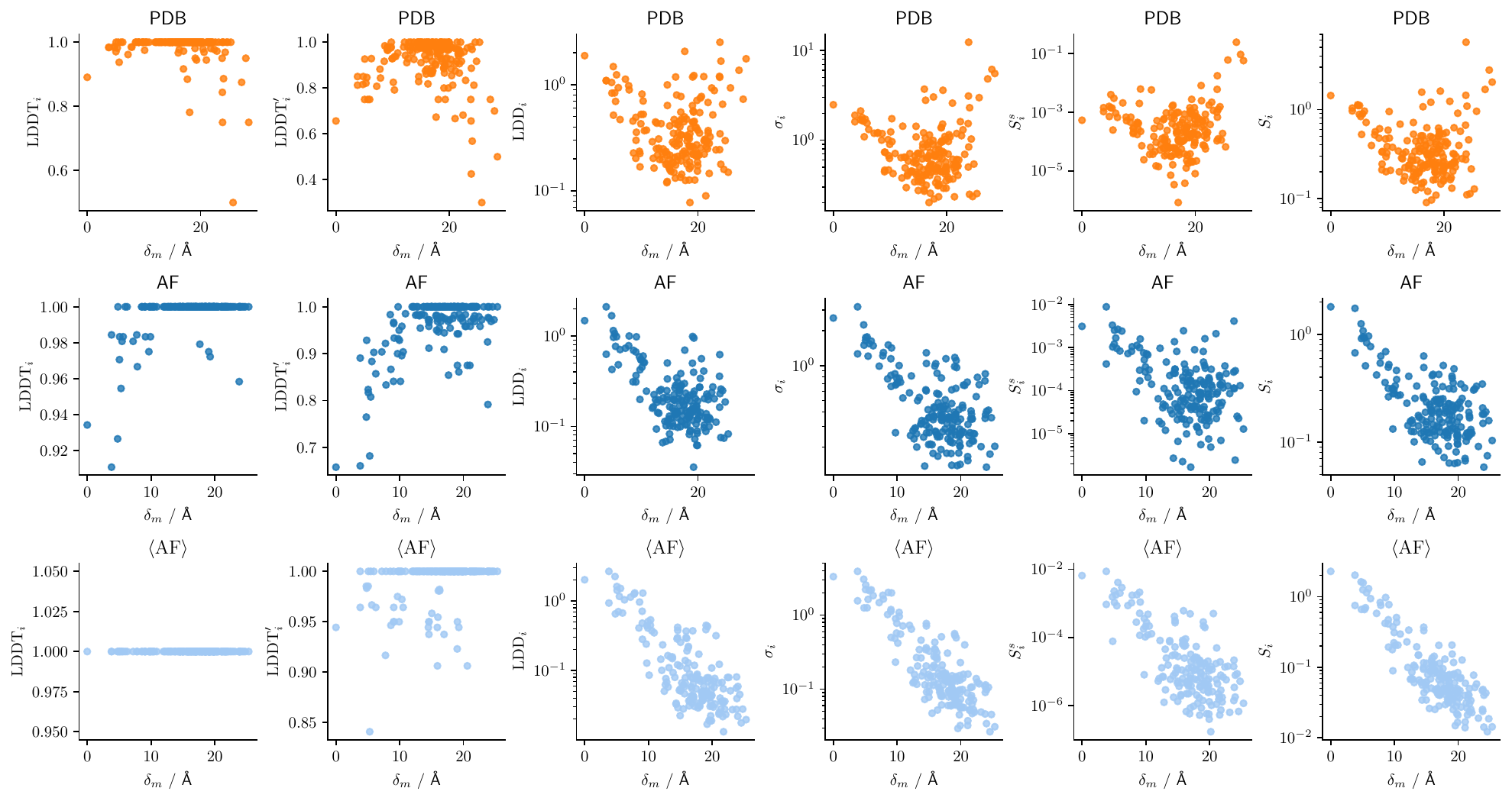}
  \caption{\label{fig:si1}
  \textbf{Examples of deformation metrics.}
  Deformation as a function of distance from nearest mutated site for deformation 
  upon mutation from WT CypA (\texttt{6U5C\_A}) to a double mutant (S99T, C115S,
  \texttt{6BTA\_A}). Deformation is shown for
  6 metrics, from left to right: LDDT; LDDT$'$ with more precise cutoffs $\zeta_k$;
  LDD; neighborhood distance, $\dNNi$; shear strain, $S^\mathrm{s}$; effective strain (ES), $\dNi$. For all metrics, $\gamma=\SI{13}{\angstrom}$.
  Deformation is calculated for PDB structures (top), AF-predicted structures (middle), and averaged AF-predicted structures ($\AFave$, bottom, see \sref{sec:ave}).
}
  \end{figure*}

  \section{Deformation metrics}
  To study the effect of mutations on structure, we need an appropriate metric
  of structural change, or \textit{deformation}. We can differentiate between measures by whether they measure local or global changes, and whether the measures are absolute, or normalized scoring functions. 
  We will explain the types of measures that have been used, and why they are not appropriate for our purpose, and discuss the pros and cons of these   alongside several other measures.

  Historically, in the field of protein structure prediction, the focus has been on measuring similarity instead of differences, and finding a well-behaved metric that can score similarity so that algorithm performance can be easily evaluated~\cite{kuf12}. 
  
  The most commonly used metric, although problematic, is the Root Mean Squared-Deviation (RMSD), which is the root-mean-square-deviation of atomic positions between a target structure (with positions $\rb_i'$) and a reference structure (positions $\rb_i$)~\cite{kuf12},
  \begin{equation}
  \label{eq:rmsd}
       \mathrm{RMSD} = \frac{1}{L} \sqrt{ \sum_{i=1}^L {\left| \rb_i' - \rb_i \right|^2}}~,
  \end{equation}
    where $L$ is the sequence length. The first step in calculating RMSD is to align the two structures via translation and rotation using the Kabsch algorithm~\cite{kabac76}. This is problematic since parts of proteins can undergo rigid-body motion, in which there is little local deformation yet the global positions undergo large-scale rearrangements. Therefore RMSD can be high in proteins that barely deform. 
  Additionally, RMSD is an absolute metric, which results in sensitivity to large deviations due to outliers (\eg, in flexible loops and tails). 
  
  RMSD is still widely used due to its simplicity, but more robust global metrics have been developed -- such as the template modelling score (TM-score)~\cite{zhapr04} and the global distance test (GDT)~\cite{zemna03} -- which align subsets of atoms rather than the whole protein, and produce scores between \SIrange{0}{1} so that effects of outliers are minimized. 
  More recently, the local distance different test (LDDT) was developed~\cite{marbi13}. LDDT is a score per residue, which compares neighbor distances in a target structure and a reference structure, and measures the fraction of corresponding distances that are within some threshold value of each other.

\begin{figure*}[t!]  \centering
  \includegraphics[width=0.99\linewidth]{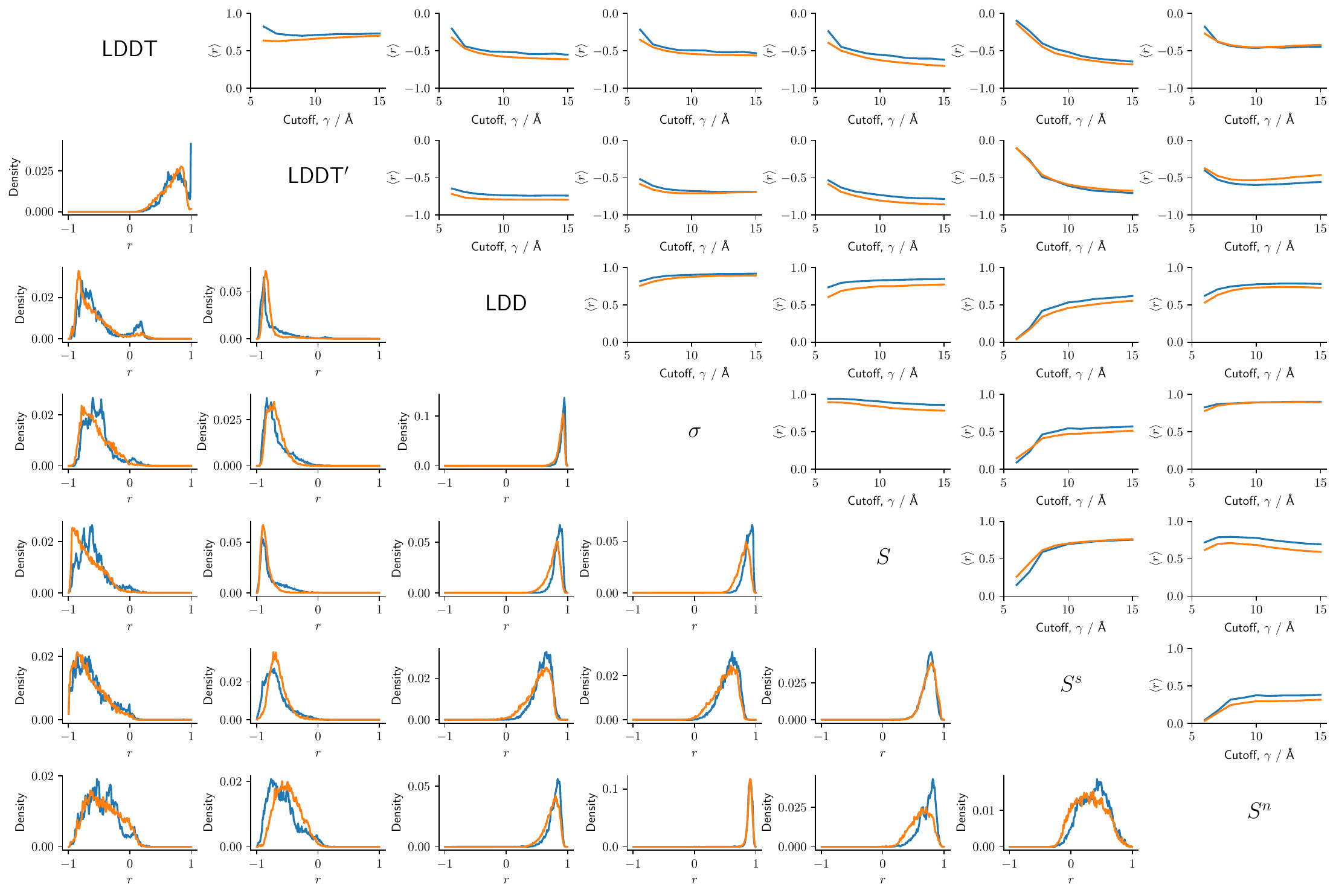}
  \caption{\label{fig:si2}
  \textbf{Correlations between deformation metrics.}
  Pearson's correlation coefficient, $r$, is calculated for each pair of proteins in our PDB dataset (\num{17813} pairs). Distribution of $r$ for each combination of metrics (Bottom, $\gamma=\SI{13}{\angstrom}$), and mean correlation $\langle r \rangle$ as a function of $\gamma$ (Top). Data is shown for both PDB (orange) and AF-predicted (blue) structures.
  }
  \end{figure*}

  Our goal is to measure local deformation, which requires a completely different type of metric. 
  First, we need a metric capable of distinguishing between deformation at different residues, since we expect mutation effects to be primarily (but not necessarily) local. This means global metrics like RMSD,   TM-score and GDT are unsuitable. Second, we want an absolute metric so that large deformation is taken for what it is, and not subject to diminishing returns.
  This implies that LDDT is not suitable, which we show in Figures \ref{fig:si1}, \ref{fig:si2}, \ref{fig:si3} and \ref{fig:si4}.
  
  In the following, we consider several metrics of local deformation that all share similar aspects,
  but differ in their origins: the local distance difference (LDD) and neighborhood distance ($\dNNi$) are mathematically related to LDDT, but result in absolute metrics rather than scores; we investigate three measures of strain (shear strain, non-affine strain, and effective strain (ES)) based on finite strain theory~\cite{lub08}.
  Note that we exclude AF-predicted residues with pLDDT $< 70$ from calculations, as we treat these as disordered residues which would always lead to high deformation.

  \subsection{Local Distance Difference Test (LDDT)}
  LDDT is a score from \SIrange{0}{1} that measures the similarity between two structures, with a value of \num{1} being maximally similar. 
  LDDT$_i$ is calculated for each residue $i$ by first defining a set $N_i$ of $n_i=|N_i|$ neighbors, $j \in N_i$, using a distance cutoff, $\gamma$.
  Residues are considered neighbors if the positions of their $\CA$ atoms, $\rb_j$, are closer than $\gamma$; \ie, if $r_{ij} = |\rb_{ij}| = |\rb_i - \rb_j| \leq \gamma$.
  For each neighbor, we calculate the distances between neighbors in both
  the reference $r_{ij}$ and the target structures $r_{ij}'$, and calculate
  the difference $\Delta r_{ij} = r_{ij}'-r_{ij}$. LDDT$_i$ is defined as the fraction of distance differences that are within a set of $k$ cutoffs $\zeta_k$,
  \begin{equation}
  \label{eq:lddt}
 \mathrm{LDDT}_i = \frac{1}{4 n_i} \sum\limits_{j \in N_i} \sum\limits_{k} \theta (\Delta r_{ij}, \zeta_k)~,     
  \end{equation}
  
  where $\theta$ is the Heaviside step function,
  \be
  \theta(\Delta r_{ij}, \zeta_k) =
    \begin{cases}
      0, & \Delta r_{ij} > \zeta_k \\
      1, & \Delta r_{ij} \leq \zeta_k~.
    \end{cases}
  \ee
  LDDT is typically calculated with $\zeta_k \in \{0.5, 1, 2, 4\}$ \AA. It is not possible to discriminate between small amounts of deformation with these cutoffs (since small deformation always gives LDDT$_i = 1$; \fref{fig:si1}),
  so we also measure a more precise alternative, LDDT$'$, using $\zeta_k \in \{0.125, 0.25, 0.5, 1\}$ \AA.

  \subsection{Local Distance Difference (LDD)}
  As an alternative to LDDT, we propose to skip the step where distances are compared with cutoff values, and instead directly compare neighbor distances in reference and target structures. 
  For each residue $i$ we calculate the local distances for each of its $n_i$ neighbors, $r_{ij} = |\rb_{ij}| = |\rb_i - \rb_j|$, and the change in the distances between the structures $\Delta r_{ij} = r_{ij}' - r_{ij}$. Then, $\mathrm{LDD}_i$ is the sum of the squared distance change,
  \begin{equation}
    \mathrm{LDD}_i = \sqrt{\sum\limits_{j \in N_i} \left( \Delta r_{ij} \right)^2}~.    
  \end{equation}
    
  One clear benefit of LDD is that does not require an alignment since it measures differences in scalar distances, yet this also means that it neglects deformation due to differences in rotations between neighbors. Beyond this, it is conceptually simple, easy to measure, and similar to LDDT.

  \begin{figure*}[t!]  \centering
  \includegraphics[width=0.99\linewidth]{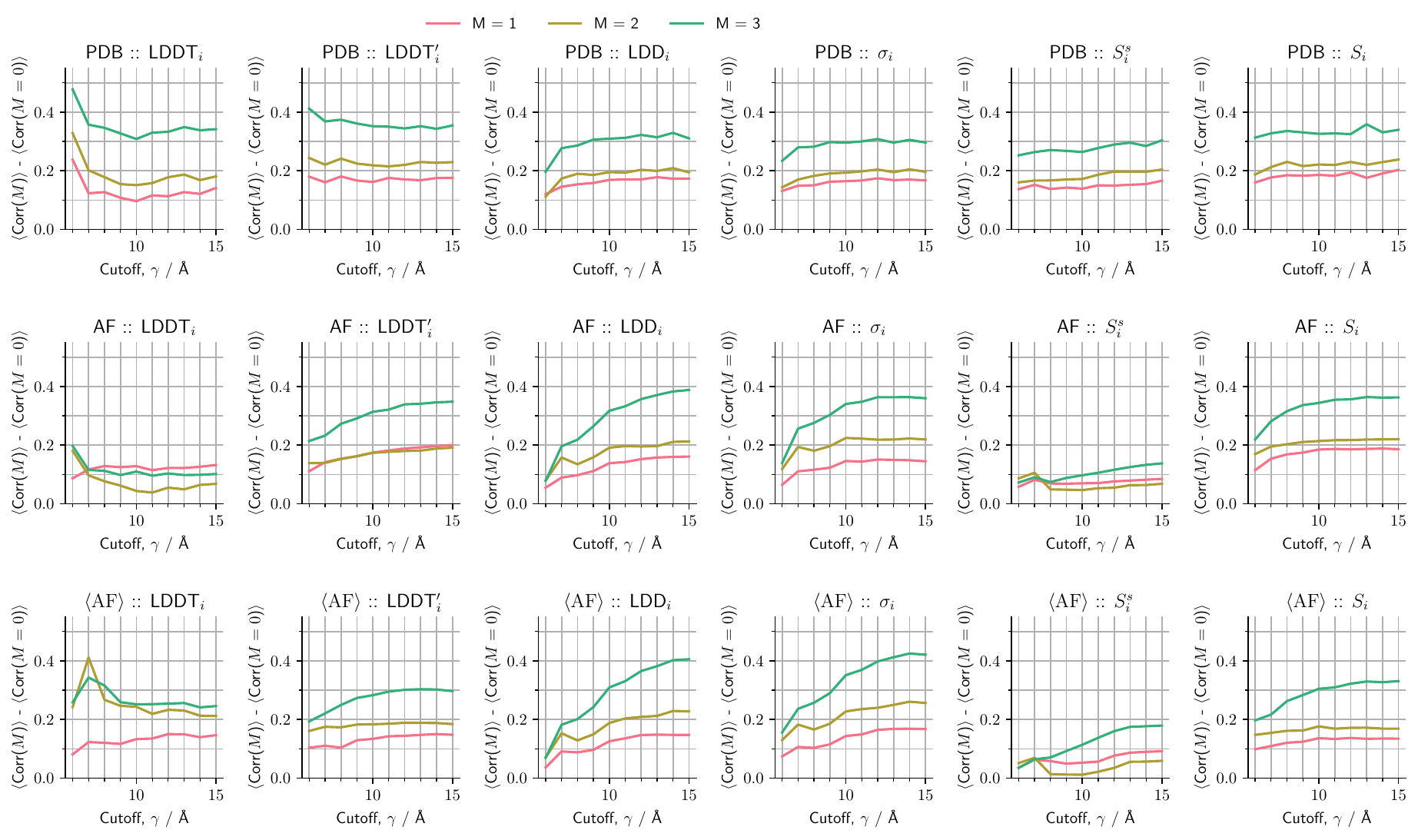}
  \caption{\label{fig:si3}
  \textbf{Correlations between PDB and PDB (top), AF (middle), and $\AFave$ deformation vectors for different metrics.}
  Residual correlation due to mutation, $\langle \corr(M) \rangle - \langle \corr(M=0) \rangle$, extracted from the distributions of correlation between PDB and AF-predicted deformation vectors, $\corr(M)$. Residual correlation is shown as a function of neighbor cutoff $\gamma$.
  Results are shown for 6 metrics, from left to right: LDDT; LDDT$'$ with more precise cutoffs $\zeta_k$; LDD; neighborhood distance, $\dNNi$; shear strain, $S^\mathrm{s}$; effective strain (ES), $\dNi$.
  For non-averaged structures (AF) we report the model generated by AlphaFold that has the highest pLDDT. We construct the average structures ($\AFave$) using all \num{5} AlphaFold models, along with \num{25} ColabFold-generated structures (\num{5} models, \num{5} repeats).
}
  \end{figure*}

  \subsection{Neighborhood Distance}
  LDDT and LDD depend only on the magnitude of the distances $\rb_{ij}$, not their orientation.
  For a more robust measure of deformation that takes into account
  deformation due to rotation of distance vectors as well as changes in their magnitude, we propose the \textit{neighborhood distance}. 
  For each residue $i$ we define a $n_i \times 3$ neighborhood tensor, $\NN_i$, as the tensor of the $\CA$ distance vectors $\rb_{ij}$ for all $j \in N_i$, where rows correspond to neighbors, and columns correspond to the distances in three dimensions, $x$, $y$, $z$.
  We get neighborhood tensors for both the reference structure, $\NN_i$, and the target structure, $\NN_i'$. 
  Then, the neighborhood distance is the norm
  of the change in distances, $\Delta \rb_{ij} = \rb_{ij}'-\rb_{ij}$, between neighborhoods.
  \begin{equation}
  \label{eq:sigmai}
  \dNNi = \left\| \NN_i - \NN_i' \right\| = 
  \sqrt{\sum_{j \in N_i}{\left| \Delta\rb_{ij} \right|^2}}~.
  \end{equation}
  This method requires that the target and reference neighborhoods are aligned.
  We achieve this by rotating (without translating) the target neighborhood using the Kabsch algorithm~\cite{kabac76}.

  We note that this metric is similar to the frame-aligned-point-error (FAPE) -- which forms part of the loss function used by DeepMind to train AlphaFold~\cite{jumna21}. $\dNNi$ and FAPE differ by the normalization factor, $n_i$, and in the frame of reference used to align the two neighborhood tensors; in AlphaFold, the reference frame is defined by the positions of the three heavy atoms (N, $\CA$, and C) in the backbone.

  \subsection{Shear Strain}
  We follow the standard treatment of finite-strain theory as implemented in \citep{eckrmp19}. To calculate strain at residue $i$, we first calculate the deformation gradient tensor, $\Fb_i$, where
  \begin{equation}
  \label{eq:affine}
      \NN_i' = \Fb_i \; \NN_i~.
  \end{equation}
  The matrix equation (\ref{eq:affine}) is usually overdetermined, so one 
  finds $\Fb_i$ using least-squares regression over the residual $\| \NN_i' - \Fb_i \; \NN_i \|^2$. We then get the Lagrangian finite strain tensor, $\Eb_i$,
  \be
  \Eb_i = \frac{1}{2}(\Fb_i\tran \Fb_i - \Ib)~,
  \ee
  where $\Ib$ is the identity matrix. Finally, as a measure of the magnitude of the shear strain, we use
  \begin{equation}
  \label{eq:shear}
        S^\mathrm{s}_i = \Tr (\Eb_i\cdot \Eb_i) - \frac{1}{3} [\Tr(\Eb_i)]^2~.
  \end{equation}

  \subsection{Non-Affine Strain}
  The non-affine strain is the deformation that is not due to any of the affine transformations -- isotropic volume expansion/contraction, or shear/twist motion. A simple estimate of the non-affine component of deformation is the residual left after solving Eq. (\ref{eq:affine}) for the affine deformation gradient tensor, $\Fb_i$,
  \begin{equation}
  \label{eq:nonaffine}
   S^\mathrm{n}_i = \norm{\NN_i' - \Fb_i \; \NN_i}^2 =
   \sum\limits_{j \in N_i}{\abs{ \rb_{ij}' - \Fb_i \;\rb_{ij} }^2}~,     
  \end{equation}
  where $j$ is the neighbor index. $S^\mathrm{n}_i$  has been previously used as an effective measure of the non-affine strain~\cite{falpr98}.
  We use a modified version of the ``atomic\_strain.py'' implementation from \citep{matscipy}.

  \begin{figure*}[t!]  \centering
  \includegraphics[width=0.99\linewidth,trim={0.5in 0.2in 0.5in 0.5in},clip]{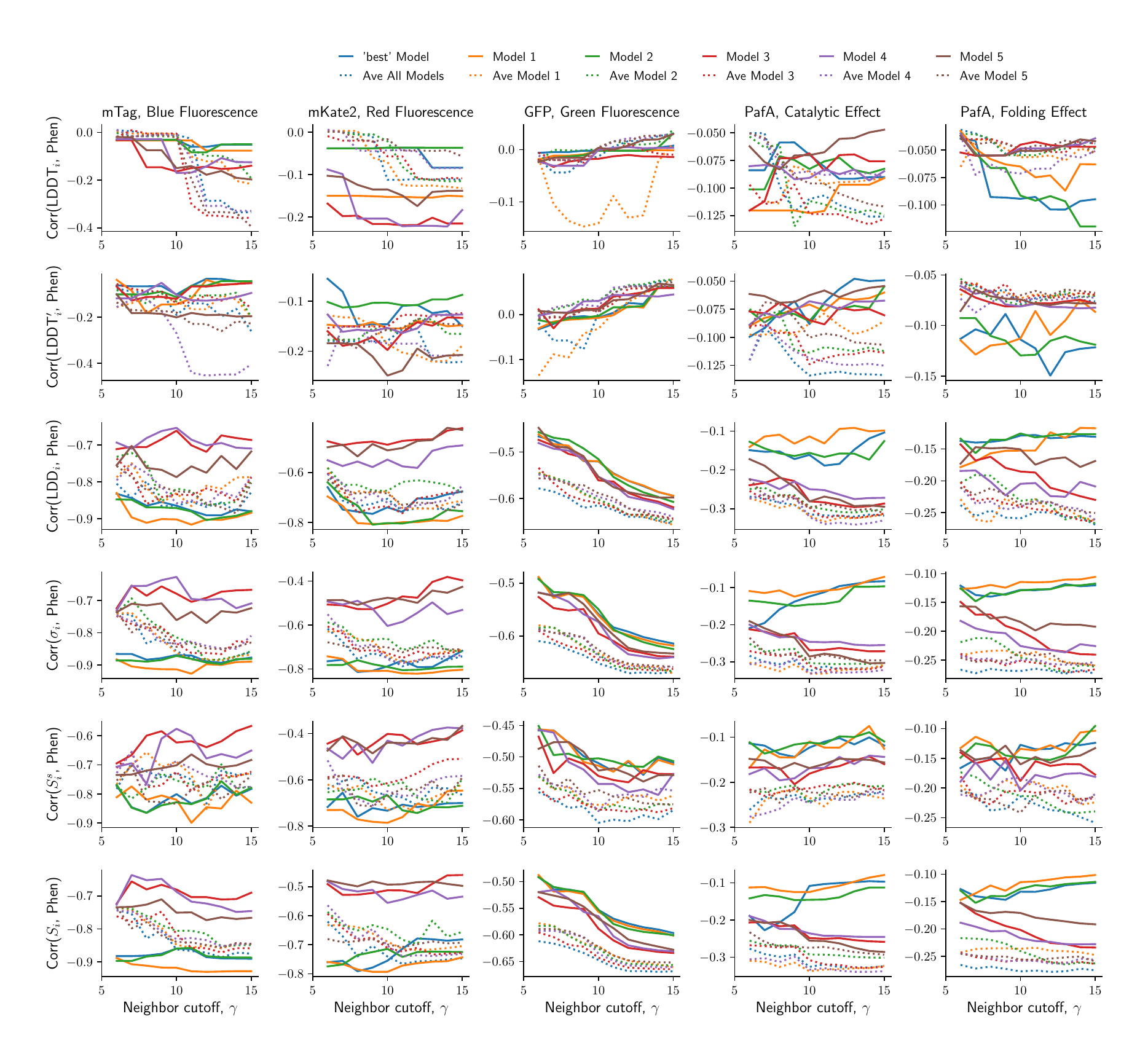}
  \caption{\label{fig:si4} 
   \textbf{Correlations between deformation metrics and phenotype measurements.}
  Correlations between deformation at residue $i$ and phenotype 
  for \num{5} sets of phenotype measurements (left to right: mTag, blue fluorescence; mTag, red fluorescence; GFP, green fluorescence; PafA, catalytic effect; PafA, folding effect), 
  for \num{6} deformation metrics
  (from top to bottom: LDDT; LDDT$'$ with more precise cutoffs $\zeta_k$;
  LDD; neighborhood distance, $\dNNi$; shear strain, $S^\mathrm{s}$; effective strain (ES), $\dNi$),
  for non-averaged and averaged structures, for all AF models.
  For each case (metric, phenotype, $\gamma$, model), we choose the residue $i$ that gives the highest correlation as a simple statistic to compare performance, and report the correlation between deformation at that specific residue with phenotype.
  The `best' model is the one with the lowest pLDDT. When averaging across structures, ``Ave All Models'' refers to averages using the entire set of structures, while the others only averaged over one out of the five AF models; an exception is PafA, where we excluded the AlphaFold-predicted models \num{1} and \num{2}, since we found that the use of a template in these cases resulted in odd predictions;
  we included the ColabFold-predicted models \num{1} and \num{2}, since we disabled the use of templates in these cases.
  }
  \end{figure*}

  \subsection{Effective Strain (ES)}
  For a more comprehensive measure of strain that does not distinguish between isotropic, shear, and non-affine components of strain, we measure the effective strain (ES); this is simply 
  the \textit{average relative change in positions},
  \begin{equation}
  \dNi = \left\langle \frac{\abs{ \Delta \rb_{ij} }}{r_{ij}} \right\rangle = \frac{1}{n_i}\sum\limits_{j \in N_i}
  \frac{\abs{ \rb_{ij} - \rb_{ij}' }}{r_{ij}}~.   
  \end{equation}
  The ES, $\dNi$, like the neighborhood distance, $\dNNi$, requires that neighborhood tensors are first aligned via rotation, and takes into account rotation-based deformation between neighbors.
  Additionally, $\dNi$ is weighted by distances of neighbors from residue $i$, so it is a dimensionless property that approximates the local strain. Since far-away residues contribute less to the total $\dNi$, this measure should be more robust to changes in the neighbor cutoff $\gamma$.

  \section{Evaluation of deformation metrics}

  \subsection{Correlations between deformation metrics}

  To probe the similarities and differences in the deformation metrics,
  we measured each metric, using cutoff values of $\gamma \in \{6, 7, ..., 16\}$ \AA,
  for both PDB and AF-predicted pairs of structures in our set of \num{17813} matched pairs.
  For each pair of structures, we calculate Pearson's correlation coefficient, $r$, between pairs of metrics,
  and report the distribution of $r$, and the average $\langle r \rangle$ as a function of $\gamma$
  (\fref{fig:si2}).

  We find that LDDT has the lowest correlation with all of the other metrics, compared to the other metrics.
  We attribute this mainly due to the low resolution of LDDT (at best \SI{0.5}{\angstrom}) compared to the average difference in backbone distances (PDB, $\SI{0.15}{\angstrom}$; AF, $\SI{0.07}{\angstrom}$, \fref{fig:si12}). Accordingly, we find that using lower LDDT cutoffs $\zeta_k$ (LDDT$'$)
  results in higher correlations with other metrics. 
  
  Importantly, we find that 
  \emph{the LDD, $\dNNi$, and effective strain (ES), $\dNi$, are all extremely well correlated}, as may be expected given they are mathematically similar. Out of the metrics related to strain, we see that the ES and non-affine strain are highly correlated, while shear strain has a much lower correlation with all of the metrics. This implies that the deformation in proteins, whether due to dynamical fluctuations ($M=0$) or due to mutations ($M>0$), has a significant non-affine component. This may not be true for large-scale deformation due to functional motion, such as that which occurs in enzymes and motor proteins.

  \subsection{PDB-AF correlations for different deformation metrics}

  In order to compare the performance of different deformation metrics for measuring mutation effects, we compare three sets of data: For PDB-PDB comparisons we identify groups of three or four PDB structures whose sequences differ by $M\in \{0, 1, 2, 3\}$ mutations. 
  For example, for $M=0$, we find three or four structures with identical sequences and calculate deformation vectors between the metrics computed for two of the pairs, and calculate Pearson's correlation coefficient, $r$.
  If there are only three available structures, then one structure is shared between the two pairs, and each pair has one unique structure; if there are four available structures, then all structures are unique. 
  We also compare sets of PDB-AF, and PDB-$\AFave$ (averaged AF-predicted structures; \sref{sec:ave}) structures. 
  
  For example, for $M=1$ we take two sequences that differ by one mutation, for which there are corresponding structures in the PDB, and calculate a deformation vector; we use AF to predict structures for the two sequences and calculate a deformation vector; we then calculate $r$ between these two vectors. 
  For average AF-predicted structures, we follow the procedure in \sref{sec:ave} to get a pair of averaged structures, calculate the deformation vector between these, and then calculate $r$ with respect to the PDB deformation vector.
  To calculate the residual correlation, $\langle \corr(M) \rangle - \langle \corr(M=0) \rangle$, we calculate the average correlation for each $M \in \{1, 2, 3\}$, and subtract this from the average correlation for all $M=0$.
  In this way, we estimate the average proportion of the correlations that is due to mutations, rather than correlated fluctuations.
  Correlations are calculated between LDDT vectors on a linear scale, since LDDT exhibits little variation. For all other metrics we calculate correlations on a log scale, since they can vary over several orders of magnitude.

  In \fref{fig:si3}, we show the residual correlation as a function of $\gamma$ for all deformation metrics, for  the three sets of data (PDB-PDB, PDB-AF, PDB-$\AFave$). 
  The highest correlations are found in the region of $10 \leq \gamma \leq 15$ \AA, for LDD, $\dNNi$, and $\dNi$. {\sbweight Spurious} high correlations are found for LDDT and LDDT$'$: most values are $\textrm{LDDT} = 1$, and for over half of proteins all residues have $\textrm{LDDT} = 1$; the high correlations reported in \fref{fig:si3} for LDDT are due to relatively few values that are $\textrm{LDDT} < 1$.

  \subsection{Deformation-Phenotype correlations for different metrics}

  In order to compare the performance of different deformation metrics for predicting the phenotypic effects of mutations, we computed correlations between phenotypic measurements and deformation at residue $i$, separately for each $i$ (\sref{sec:phendata}, \sref{sec:phencorr}). 
  In \fref{fig:si4} we compare performance for: all \num{5} phenotypes, \num{6} deformation metrics, different AF models, and either non-averaged or averaged structures; we show results for deformation at whichever residue $i$ has the highest correlation with phenotype.
  With regards to the different deformation metrics, we find that 
  \emph{the metrics that show the highest correlations between PDB-AF structures (LDD, $\dNNi$, and $\dNi$), also show the highest deformation-phenotype correlations.} 
  LDDT results in especially poor correlations with phenotype, and using the more precise distance cutoffs $\zeta_k$ (LDDT$'$) does not help. Shear strain $S^\mathrm{s}_i$ is only slightly worse than effective strain $\dNi$.

  \begin{figure}[t!]  \centering
  \includegraphics[width=0.99\linewidth]{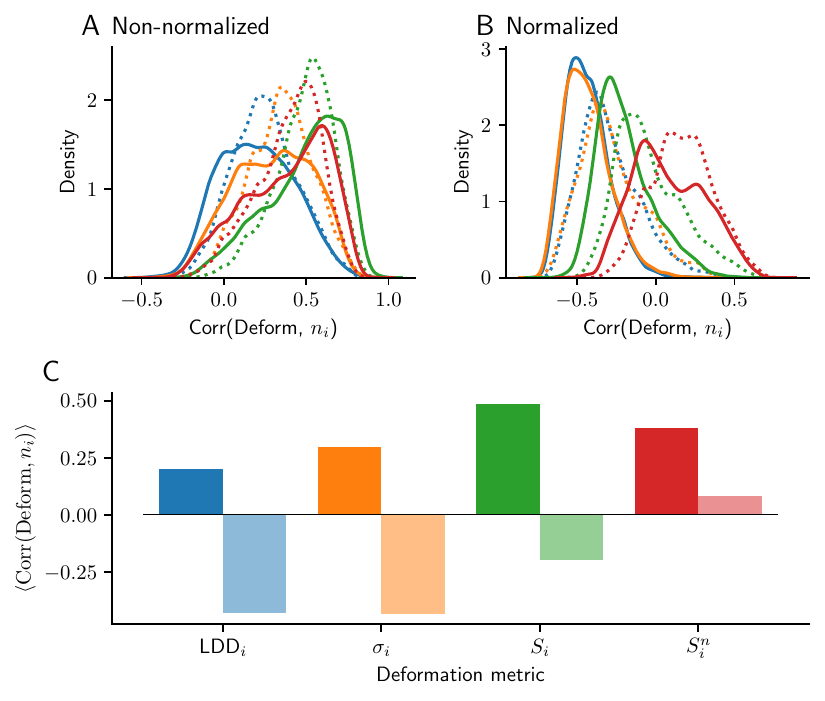}
  \caption{\label{fig:si6c}
  \textbf{Effect of normalization.}
  A-B: Distribution of correlations between deformation (LDD, blue; neighborhood distance, orange; ES, green; non-affine strain, red) and number of neighbors $n_i$, over all matched pairs of proteins ($0 \leq M \leq 3$) for PDB (solid line) and AF-predicted structures (dotted line). Distributions are shown for non-normalized metrics (A) and normalized metrics (B).
  C: Average correlation for different metrics. Results are shown for non-normalized metrics in bold and for normalized metrics in pastel color.
  }
  \end{figure}

  \subsection{Normalizing by number of neighbors}
  In our definitions of $\textrm{LDD}_i$, $\sigma_i$ and $S_i$, there is the choice to normalize by the number of neighbors, $n_i$, which gives us the average deformation per neighbor. There is no inherently correct choice, as the two types of metric (normalized or non-normalized) simply report different information: Normalizing results in a metric that describes mean deformation, while not normalizing results in a metric that also takes into account the number of neighbors. Thus, one might suspect that non-normalized metrics are correlated with the number of neighbors. We find this to be true, however we also find that normalized metrics are anti-correlated with the number of neighbors (\fref{fig:si6c}). We attribute this to the increased flexibility and variance across repeat predictions that we observe in surface residues that have fewer neighbors.
  
  It seems that due to these opposing effects -- few neighbors leads to higher flexibility and higher mean deformation, and many neighbors leads to higher overall deformation due to summing over many neighbors -- there is no clear overall effect of normalization. Importantly, we find that there is no effect of normalization on deformation-phenotype correlations. For PDB-AF deformation correlations, we find that normalizing results in higher correlations when the deformation metric uses the L1 norm (\eg, $S_i$) instead of the L2 norm. We find that not normalizing results in higher correlations when the deformation metric uses the L2 norm instead of the L1 norm. The normalized and un-normalized versions of each metric also correlate highly with each other. Ultimately, we find that the results are not particularly sensitive to the exact form of the deformation metric, and exclusively report ES in the main manuscript.

  \begin{figure*}[hp!]  \centering
  \includegraphics[width=0.99\linewidth]{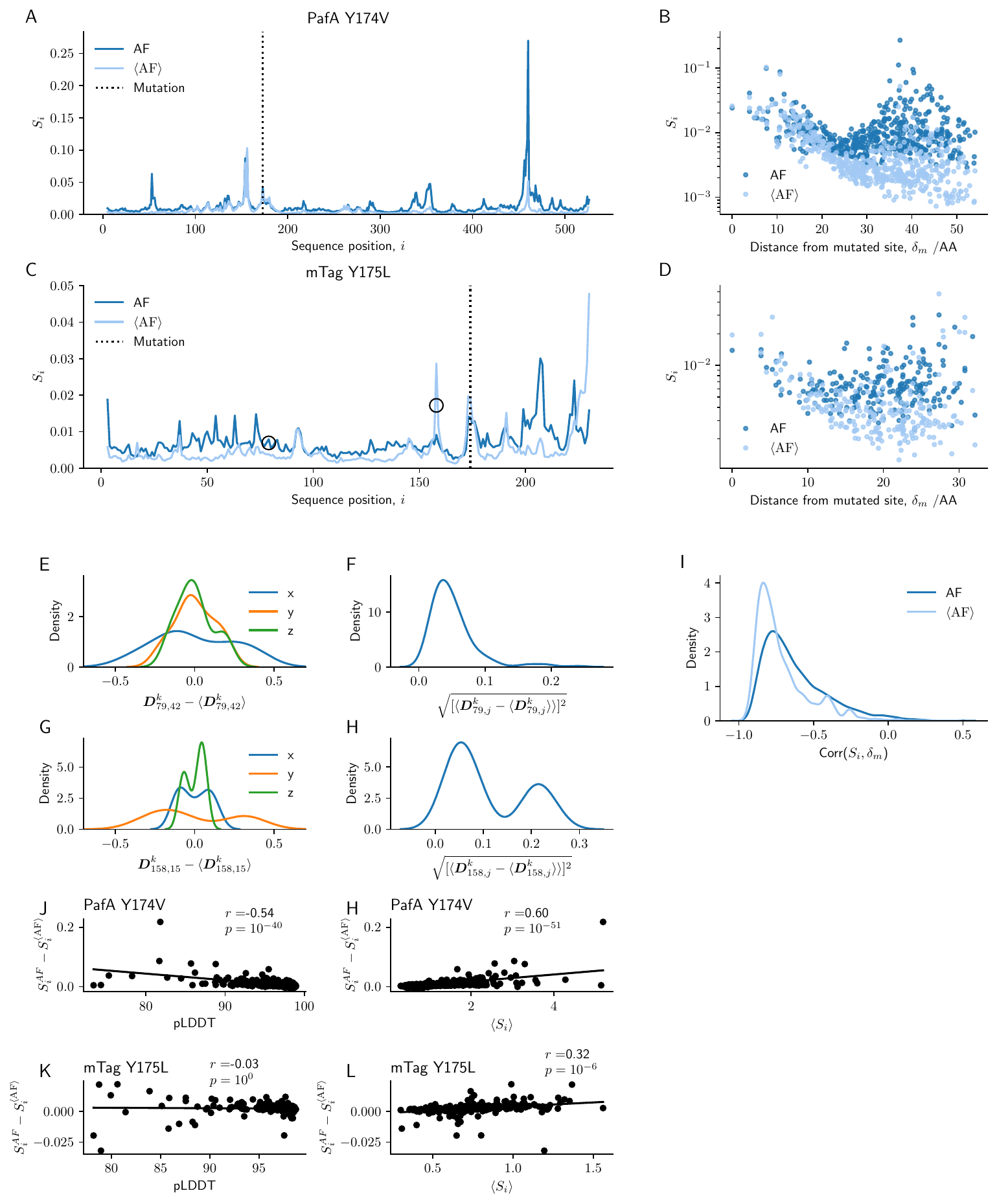}
  \caption{\label{fig:si5}
  \textbf{Averaging across multiple AF predictions.}
  A-D: Examples of strain per residue $\dNi$ for PafA (A: WT vs Y174V) and mTag (C: WT vs I175L) calculated using single AF structures (AF) and averaged structures ($\AFave$).
  Dotted line indicates the mutated residue; circles indicate residues 79 and 158 (C, mTag). Effective strain (ES) per residue against distance from the mutated site, $\delta_{m,i}$, for each residue $i$ for PafA (B) and mTag (D).
  E-H: Distributions of $x$, $y$ and $z$ components of the (neighborhood-aligned) vectors $\NN_{79,42}^k$ (E) and $\NN_{158,15}^k$ (G) across all repeat predictions $k$ of the mTag WT and I175L, normalized by the average $\langle \NN_{ij}^k\rangle$ across $k$.
  Distribution of standard deviations of elements ($\delta x_{i,j}$, $\delta y_{i,j}$ and $\delta z_{i,j}$) in aligned neighborhood tensors across all repeat predictions of the mTag WT and I175L for residues 79 (F) and 158 (H).
  I: Distribution of correlations between strain per residue $\dNi$ and
  distance from the nearest mutated site $\delta_{m,i}$ (I) for all pairs of 
  AF-predicted structures (`best' model) in our PDB dataset that are mutated ($M>0$).
  J-L: Difference in deformation between AF and $\AFave$ structures as a function of pLDDT (J, K) and repeat-prediction variability $\langle \dNi \rangle$ (H, L), for PafA WT vs variant Y174V (J,H) and mTag WT vs variant Y175L (K, L).
  }
  \end{figure*}

  \section{Averaging AF-predicted structures}\label{sec:ave}
  We observed that deformation between repeat measurements (structures with same sequence, $M=0$), whether PDB or AF, tends to depend on local flexibility (main paper, App. 2B). 
  Thus, we consider that it may be possible to smooth out these structural fluctuations by averaging over many AF predictions.
  We take into account the fact that proteins can undergo large-scale, rigid-body transformations, and thus do not average over the entire \emph{global} protein structure. 
  This would almost certainly lead to extreme, unphysical average configurations. Instead, we average over \emph{local} neighborhoods $\NN_i$.
  For each residue $i$, we extract the local neighborhoods per repeat prediction $k$, $\NN_i^k$, using a distance cutoff $\gamma$. 
  We choose one neighborhood $k_0$ as a reference neighborhood.
  We then remove neighbors that are not contained in all $k$ neighborhoods, such that the neighborhood is defined by the neighbors that are within a distance of $\gamma$ from residue $i$ in all predicted structures $k$. 
  We then rotate all neighborhoods $k \neq k_0$ with respect to neighborhood $k_0$ using the Kabsch algrithm~\cite{kabac76}, and average over neighborhood $k_0$ and the rotated neighborhoods $k \neq k_0$ (a total of $n_k$ neighborhoods) to get an average neighborhood $\langle \NN_i \rangle$,
  \be
  \langle \NN_i \rangle = \frac{1}{n_k} \sum\limits_{k} \NN_i^{k*} ~,
  \ee 
  where $\NN_i^{k*}$ is the neighborhood of residue $i$ from repeat prediction $k$ after rotation.
  We refer to these averaged structures (sets of $\langle \NN_i \rangle$) using the notation $\AFave$.

  We note that averaging in this way can lead to odd configurations, that can only be accessed through bond angles that are unphysical. Thus, these averaged configurations should not be used to study structure. 
  Instead, this is a viable approach to studying differences in structure, as the effect of averaging is to get a better estimate of a summary statistic (the `average structure'), which can be used to measure small differences.

  To illustrate how averaging can reduce the strain that arises from fluctuations rather than mutations (also see \fref{fig:si1}), we show deformation against sequence position for two examples (PafA WT vs Y174V, and mTag WT vs I175L; \fref{fig:si5}A-D). 
  In \fref{fig:si5}A, there are many high-strain regions that are far away from the mutated site when using the non-averaged structures (\fref{fig:si5}B). 
  Averaging over neighborhoods reduces the strain in these areas and reveals that strain is much more dependent on the distance from the mutated site (\fref{fig:si5}B).
  To show how this effect generalizes, we calculate the correlation between strain and distance from the nearest mutated site for all mutated protein pairs in the PDB dataset (\fref{fig:si5}I), and find that averaging tends to increase this correlation.
  This indicates that 
  \emph{averaging is able to smooth out the effects of non-mutation fluctuations}.

  We also note that it is possible for averaging to result in increased strain, as can be seen for residue \num{158} in \fref{fig:si5}C. These cases can arise due to the fact that conformations can exist in discrete populations (\eg, rotamers).
  To illustrate this, we investigate the neighborhood of residue \num{158} in more detail, by comparing neighborhoods from different predicted structures.
  We first rotate all neighborhoods $\NN_{158}^k$ to an arbitrary reference neighborhood $k_0$. This allows us to examine the variance of individual components of the neighborhoods, and how they differ across predictions. 
  We see that the $x$, $y$ and $z$ components of $\NN_{158}^k$ corresponding to neighbor $j=15$ and different repeat predictions $k$ exhibit bimodal distributions (\fref{fig:si5}G).
  By looking at the distribution of the variances of each component (each neighbor $j$, and each spatial dimension), we see that this is quite common -- the rightmost peak in \fref{fig:si5}H corresponds to components with bimodal distributions. This stands in contrast to similar results for the neighborhood of residue \num{79}, where there are some bimodal distributions amongst the components of the neighborhood tensor (\fref{fig:si5}E), but there are far fewer of these overall (\fref{fig:si5}F).
  We have thus learned that due to the presence of discrete local conformations, averaging can fail to reduce the strain across structures if too few repeats are used, but in general this is a promising strategy to achieve more precise predictions of mutation effects.

  Since some proteins (and some parts of proteins) are more flexible than others, it is useful to have an estimate of how many structures one would need to average over to smooth out the non-mutation fluctuations. 
  AlphaFold's predicted confidence score, pLDDT, is a suitable candidate for this. Using the above examples (\fref{fig:si5}A,C), we compare pLDDT with the change in deformation after averaging, $\dNi^\AF - \dNi^\AFave$ (\fref{fig:si5}J,K). We find that pLDDT is only correlated
  with the change in deformation for one out of two cases. 
  We suspect that this is due to pLDDT being a low-resolution metric, since most residues have similar pLDDT scores.
  An alternate approach is to predict many repeat structures, and to calculate the variance between $n_k$ repeat measurements of deformation, $\langle \dNi \rangle$,
  \be\label{eq:repvar}
  \langle \dNi \rangle = \frac{1}{n_k}\sum\limits_{k} \dNi^k~.
  \ee
  This is a more direct measurement of the variance in repeat predictions, and it is a much better predictor of the change in deformation due to averaging. The repeat-prediction variance can thus be used to guide decisions about how many structures one should average
  over to get a good estimate of mutation effects. If the repeat-prediction variance is low, then few repeats are sufficient; if the repeat-prediction variance is very high, then perhaps AlphaFold will not be useful at predicting mutation effects in that region.

  \begin{figure*}[b]  \centering
  \includegraphics[width=0.99\linewidth,trim={0.5in 0.2in 0.5in 0.5in},clip]{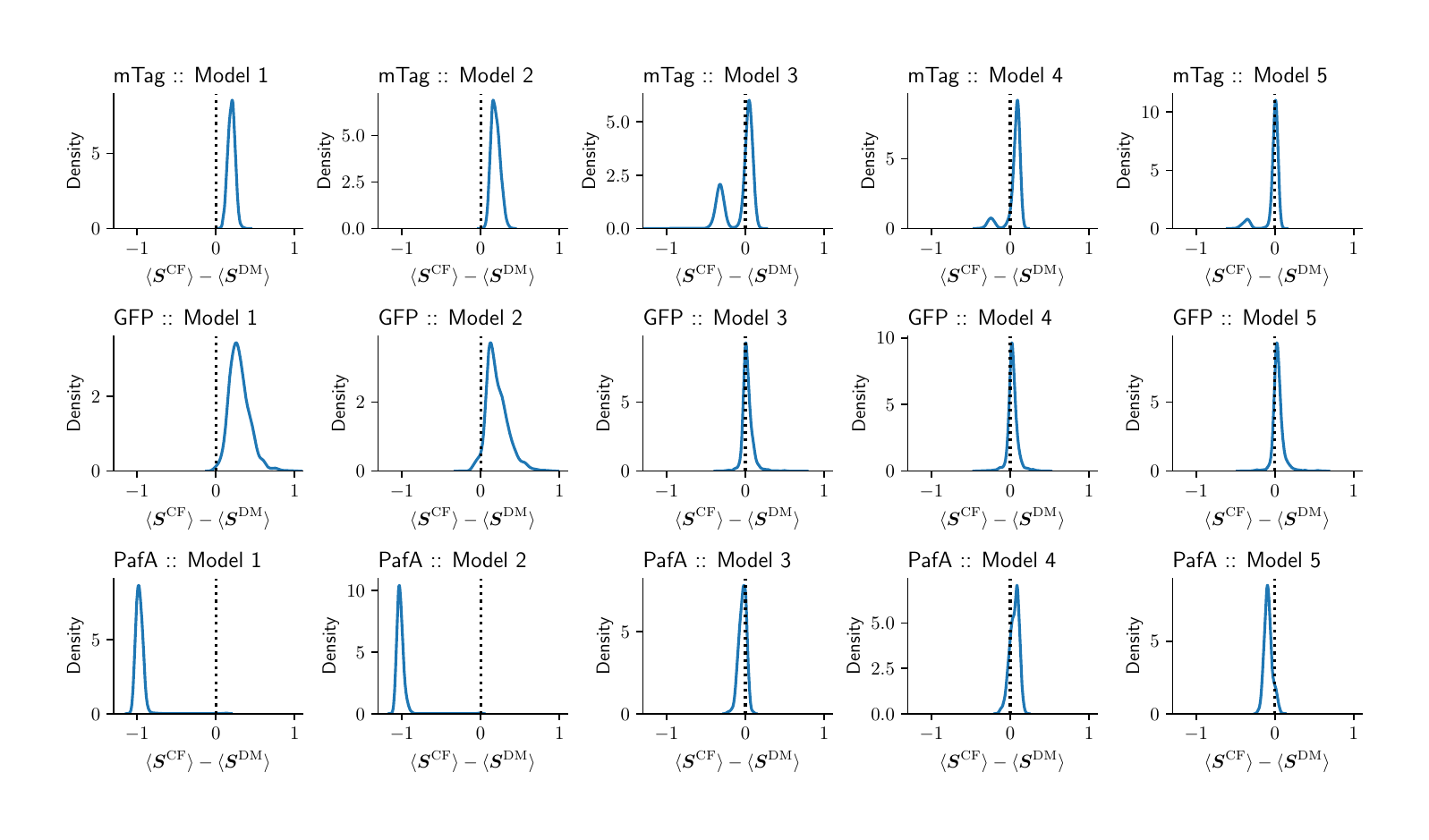}
  \caption{\label{fig:si6}
  \textbf{Comparing DeepMind and ColabFold implementations of AF.}
  Difference in the average effective strain (ES) (with $\gamma=\SI{13}{\angstrom}$ ) per pair of WT and mutant structure, $\langle \dNi \rangle$, for structures predicted using both DeepMind, $\langle \dNi^\mathrm{DM} \rangle$, and ColabFold, $\langle \dNi^\mathrm{CF} \rangle$, implementations. Results are shown for each high-throughput dataset (mTag, GFP, PafA), and for each AF model.
  }
  \end{figure*}

  \subsection{Effect of averaging on PDB-AF correlations}

  When we compare the PDB-AF correlations with PDB-$\AFave$ correlations (\fref{fig:si3}), we typically no difference for the best deformation metrics (LDD, $\dNNi$, and $S_{e}$).
  We attribute this to the high levels of repeat-measurement deformation in PDB structures; increasing precision of AF predictions cannot increase correlations since they are limited by PDB imprecision.

  \subsection{Effect of averaging on AF-phenotype correlations}

  To understand whether averaging structures can produce more accurate predictions, we also look at the effect of averaging on deformation-phenotype correlations. 
  In \fref{fig:si4} we compare performance for: all \num{5} phenotypes, \num{6} deformation metrics, different AF models, and either non-averaged or averaged structures. 
  In the majority of cases, averaging results in correlations that are stronger by about  $| \Delta r |  = \numrange{0.05}{0.1}$. 
  Excluding the results for LDDT (which performs poorly in general), averaging across all models tends to produce the best results (``Ave All Models'', blue dotted line), although there is often little difference between the results obtained for averaging over specific models (other dotted lines). 
  In comparison, for non-averaged structures, we see much higher variability in performance depending on the AF model used. This suggests that averaging across many models is a useful way to measure mutation effects, and can be used instead of simply choosing the model with the highest pLDDT.

  The results for the protein mTag (blue and red fluorescence) are an exception.
  They show that for two models, \num{1} and \num{2} (in this case the `best' model almost always corresponds to either of these two models), the non-averaged structures perform better than any of the averaged structures. 
  We investigated why this may be the case, and found that it is due to differences in the variability of predictions using DeepMind's AlphaFold versus ColabFold.
  We calculate the average effective strain (ES) per pair of WT and mutant structures for each high-throughput dataset (mTag, GFP, PafA), for both AlphaFold implementations,
  \be
  \langle \dNv \rangle = \frac{1}{L'} \sum \limits_{i} \dNi ~,
  \ee
  where $L'$ is the length of the protein, minus the number of disordered residues (\ie, where pLDDT < 70). 
  We then calculate the difference in the average ES between implementations, and show the distribution for each AF model in \fref{fig:si6}. 
  In the majority of cases, the DeepMind implementation produced lower average strain. The major exception to this is that DeepMind models 1 and 2 produced extremely high strain for PafA, which we consider an anomalous result that arises due to the use of structural templates in prediction (\sref{sec:model}).
  We also see that DeepMind models 3, 4 and 5 occasionally produced high strain for mTag variants. 
  
\begin{figure}  \centering
  \includegraphics[width=0.99\linewidth,trim={0.5in 0.2in 0.5in 0.5in},clip]{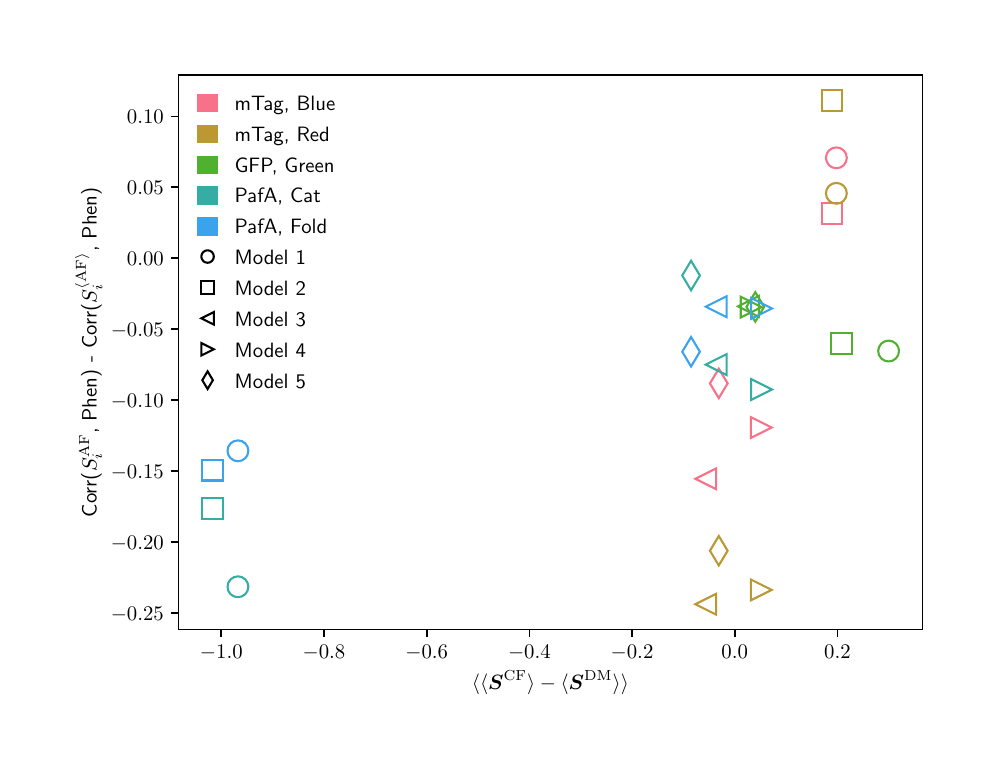}
  \caption{\label{fig:si7}
  \textbf{Differences between AF implementations correlate with differences in AF vs $\AFave$ accuracy in predicting phenotype.}
  Average difference in the average strain per pair of WT and mutant structure,
  $\langle S_{e} \rangle$, for DeepMind (DM) and ColabFold (CF) implementations, against the difference in correlation between effective strain (ES) (with $\gamma=13$ \AA)
  and phenotype (blue fluorescence, red fluorescence, green fluorescence, catalytic effect,
  folding effect), for all AF models. Pearsons's correlation coefficient is calculated for
  the full \num{25} points ($r = 0.53$, $p=0.006$), and also for the subset of points
  excluding the four points on the left ($r=0.51$, $p=0.019$).
  }
  \end{figure}

  We then compared the average differences between implementations, ColabFold and DeepMind, $\langle \langle \dNv^\mathrm{CF} \rangle - \langle \dNv^\mathrm{DM} \rangle \rangle$, for each model, to the differences between strain-phenotype correlations calculated using non-averaged (AF) and averaged ($\AFave$) structures. 
  We find a significant correlation between these differences (\fref{fig:si7}), which suggests that the cases where averaging produced worse correlations (mKate2 and mTagBFP2, models \num{1} and \num{2}) were due to the higher variability of predictions using ColabFold (since ColabFold structures constitute the majority of the structures used in averaging).
  We conclude that, in general, averaging structures appears to produce better results, but it depends on the quality of the prediction. When running ColabFold we used \num{6} recycles per structure, but this might be insufficient for mTag.

  \section{Range of mutation effects}

\begin{figure}[t!]  \centering
  \includegraphics[width=0.99\linewidth]{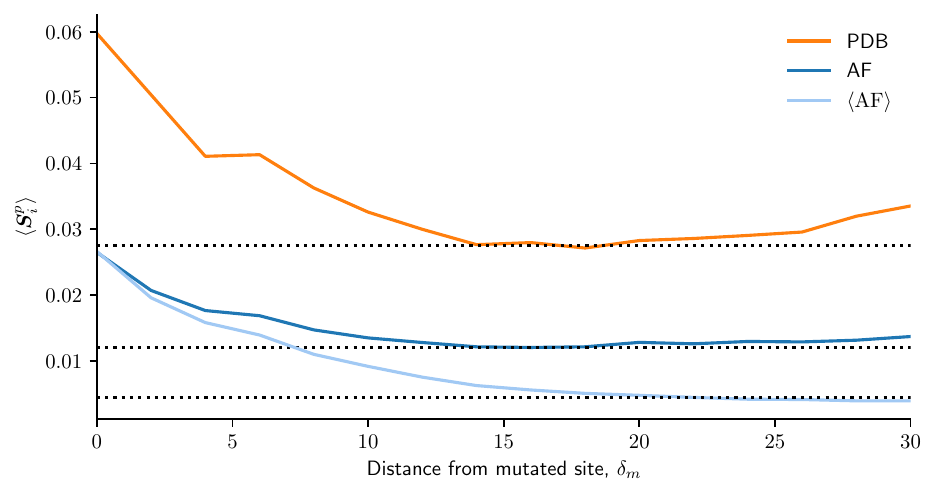}
  \caption{\label{fig:si8}
  \textbf{Average range of mutation effects.}
  Average deformation, $\langle \dNv^p_i \rangle$, across all protein pairs $p$ and all residues $i$ within a certain distance from the nearest mutated site $\dm$, as a function of 
  $\dm$, calculated respectively using PDB, AF and $\AFave$ pairs. Bins of size \SI{2}{\AA} were used. Dotted lines are visual guides.
  }
  \end{figure}

  We calculate the average range of mutation effects by looking at deformation
  as a function of distance from the nearest mutated site, $\dm$.
  We calculated the average ES, $\langle \dNv^p_i \rangle$,
  across all mutated ($M>0$) pairs of proteins $p$, and all residues $i$
  within $\dm$ bins of size \SI{2}{\AA} (\fref{fig:si8}). When we consider the
  full set of PDB pairs, we find that deformation reaches a plateau at about
  \SI{14}{\AA}; for AF pairs, the range appears to be \SI{16}{\AA}, and
  for $\AFave$ pairs the range is about \SI{18}{\AA}. Note that while we show allosteric effects in \fref{fig:si8} up to ranges of almost \SI{2}{nm}, these represent average ranges of mutation effects;
  this does not preclude the possibility that there are much longer-range allosteric effects due to mutations.

  \section{When do AF predictions correlate with PDB data?}
  To understand why some PDB-AF correlations are high and others low,
  we looked at three \textit{factors}: (i) protein flexibility, (ii) secondary structure,
  and (iii) effect size of the mutation. For each type of factor, we studied
  multiple scalar \textit{properties}, listed below. For each property,
  we measured Pearson's correlation coefficient, $r$, between the property and 
  the corresponding PDB-AF correlation for a set of protein pairs. We used
  the non-redundant set, and resampled \num{1000} times to get
  distributions of $r$ to account for sampling variance.

  \begin{figure}[t!]  \centering
  \includegraphics[width=0.99\linewidth]{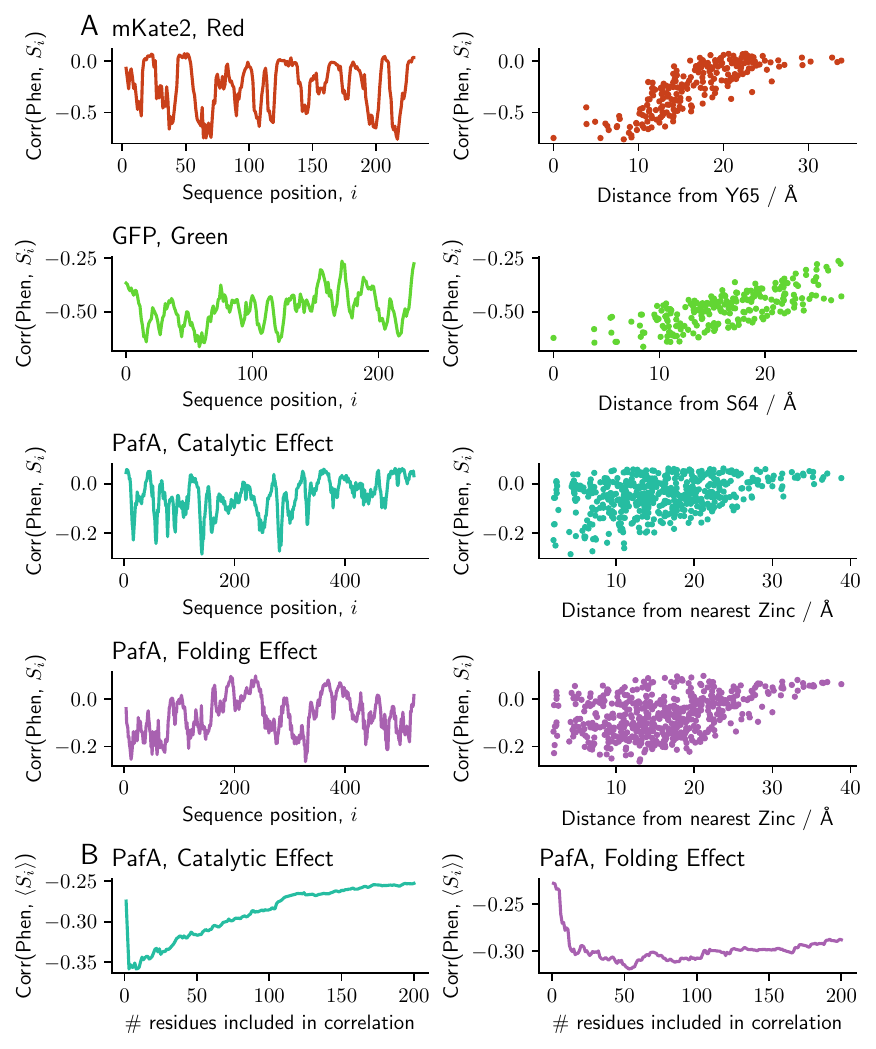}
  \caption{\label{fig:si8b}
  A: (Left) Correlation between deformation at residue $i$ and phenotype, for all sequence positions $i$: mKate2, red fluorescence; GFP, green fluorescence; PafA, catalytic effect; PafA, folding effect. (Right) Correlation between deformation at residue $i$ and phenotype as a function of distance from functional residues / co-factors: mKate2, residue Y65; GFP, residue S64; PafA, distance from the nearest Zinc co-factor.
  B: Correlation between phenotype and the mean deformation at the set of residues $i$ that (individually) correlate best with the phenotype, as a function of the size of the set.
  }
  \end{figure}

  \begin{itemize}[leftmargin=*]
    \item Protein flexibility
    \begin{itemize}[leftmargin=*]
      \item \textbf{B-factor:} B-factors are obtained from X-ray scattering experiments
            for each atom. The B-factor is related to the uncertainty in the position of an atom's
            coordinates, and is known to correlate with local flexbility~\cite{suncr19}.
      \item \textbf{pLDDT:} pLDDT is predicted by AlphaFold, and is supposed to predict
            the uncertainty of the structure predictions of individual residues, much like how LDDT
            scores the accuracy of structure predictions of individual residues.
            Recent studies have indiciated that it negatively correlates with local flexibility~\cite{guosr22,maps23}.
      \item \textbf{RSA:} Relative solvent accessibility (RSA) of a protein residue is the
            amount of an amino acid's surface that is accessible to solvent, compared to the amino acid's total surface area.
            Residues with higher RSA are less sterically constrained by other residues, and thus tend to be more flexible~\cite{zhaps09}.
      \item \textbf{Prediction Variance:} We measure the mean deformation per residue by comparing
            multiple predictions of the same protein sequence, $\langle \dNi \rangle$ (\eref{eq:repvar}). AlphaFold is a stochastic algorithm,
            which has been shown to sample some of the conformational diversity of real proteins~\cite{delel22,salbi22}.
            Thus, we expect that the degree of deformation per residue across repeat predictions should be 
            correlated with the degree of conformational flexibility.
    \end{itemize}
    \item Secondary Structure
    \begin{itemize}[leftmargin=*]
      \item \textbf{$\alpha-$Helix:} We calculate the fraction of residues in $\alpha$-helices per protein~\cite{kabbi83}.
      \item \textbf{$\beta-$Strand:} We calculate the fraction of residues in $\beta$-strands per protein~\cite{kabbi83}.
    \end{itemize}
    \item Magnitude of mutation effect
    \begin{itemize}[leftmargin=*]
      \item \textbf{PDB-PDB Correlation:} Some protein pairs tend to have more reliable correlations
            between their deformation vector, and deformation vectors of other matched pairs.
            This could be because the mutations have strong effects, which can be reliably measured.
            Thus we look at the subset of protein pairs for which we can measure the average PDB-PDB correlations
            with other matched pairs, and see whether this correlates with the PDB-AF correlations.
            We expect that high PDB-PDB correlations will be predictive of high PDB-AF correlations.
      \item \textbf{Deformation Magnitude:} We calculate the magnitude of the deformation at the
            mutated site, $S_\mathrm{m}$, for both PDB and AF structures. High $S_\mathrm{m}$ indicates
            a large mutation effect.
      \item \textbf{BLOSUM Score:} BLOSUM62 scores describe how likely a particular amino
            acid substitution is, and are calculated from the frequency of substitutions in sequence alignments~\cite{henpn92}.
            Low BLOSUM scores indicate non-conservative mutations, and are expected to lead to
            larger mutation effects.
      \item \textbf{MSA mut-freq:} A more direct way of measuring the compatibility of certain mutations at specific positions in a protein sequence is to measure the frequency of a mutation in an MSA at the mutated position. One might expect that high-frequency mutations would lead to small changes, and thus may have lower correlations. We actually find the opposite effect, as high-frequency mutations have higher PDB-AF correlations. This may indicate that there is more information in the MSA, which could improve the prediction. Explicit effects of MSA size and coverage ought to be examined in a future study.
    \end{itemize}
  \end{itemize}

  \section{Deformation-phenotype correlations}\label{sec:phencorr}
  \subsection{mKate2}
  We calculate deformation (\ie~the ES, $\dNi$) of all variants with respect to WT mKate2, at all
  sequence positions $i$. We then calculate the correlation of $\dNi$
  at each position $i$ with red fluorescence. Many positions have strong correlations
  with fluorescence, and the degree of correlation depends on how close
  each residue is to Y65, which is a site for covalent binding to the chromophore (\fref{fig:si8b}A).
  Correlations are robust to choice of metric (\fref{fig:si4}).

  \subsection{GFP}
  We calculate deformation of all variants with respect to WT GFP, at all
  sequence positions $i$. We then calculate the correlation of $\dNi$ with green fluorescence. 
  Many positions have strong correlations with fluorescence, and the degree of correlation depends on how close
  each residue is to S64, which is a site for covalent binding to the chromophore (\fref{fig:si8b}A).
  Correlations are robust to the choice of metric (\fref{fig:si4}).

  \begin{figure}[t!]  \centering
  \includegraphics[width=0.90\linewidth]{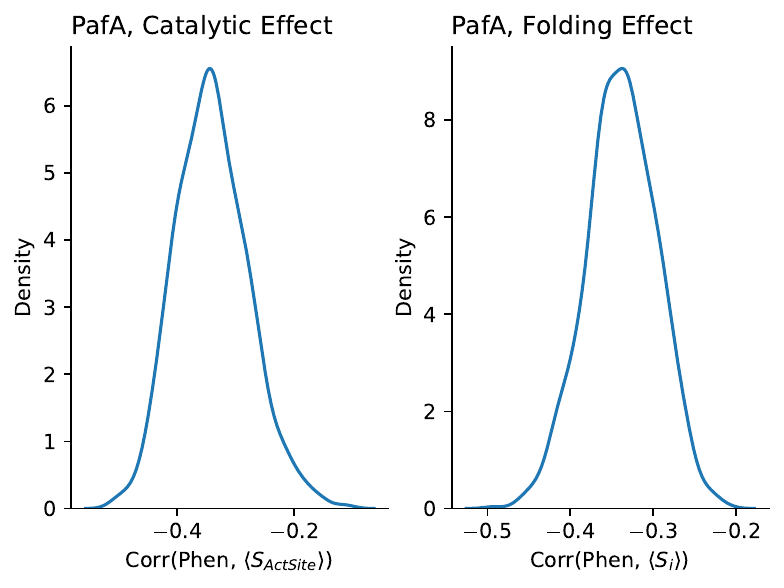}
  \caption{\label{fig:si8c} 
  Distribution of Pearson's correlation between phenotype and deformation
  obtained by randomly sub-sampling with replacement. Sub-sample size is half of the original sample size; \num{1000} sub-samples were drawn.
  }
  \end{figure}

  \subsection{PafA}
  We calculate the deformation $\dNi$ of all variants with respect to WT PafA, at all
  sequence positions $i$. We then calculate the correlation of $\dNi$ with the measured catalytic effect, and folding effect.
  Many positions are correlated with either the folding effect and/or catalytic effect (\fref{fig:si8b}A).
  For the catalytic effect, the strongest correlations are close to the binding site (which contains zinc atoms), while for the folding effect, the strongest correlations are more dispersed.
  We find higher correlations between deformation and phenotype when we take the mean 
  of deformation values amongst the top $\lambda$ residues that correlate best with the phenotype (\fref{fig:si8b}B).
  In the case of the catalytic effect, we take the first $\lambda = 5$ residues,
  while for the folding effect, we take $\lambda=50$.
  Correlations are robust to the choice of metric (\fref{fig:si4})
  and robust to subsampling (\fref{fig:si8c}).

  \begin{figure}[t!]  \centering
  \includegraphics[width=0.99\linewidth]{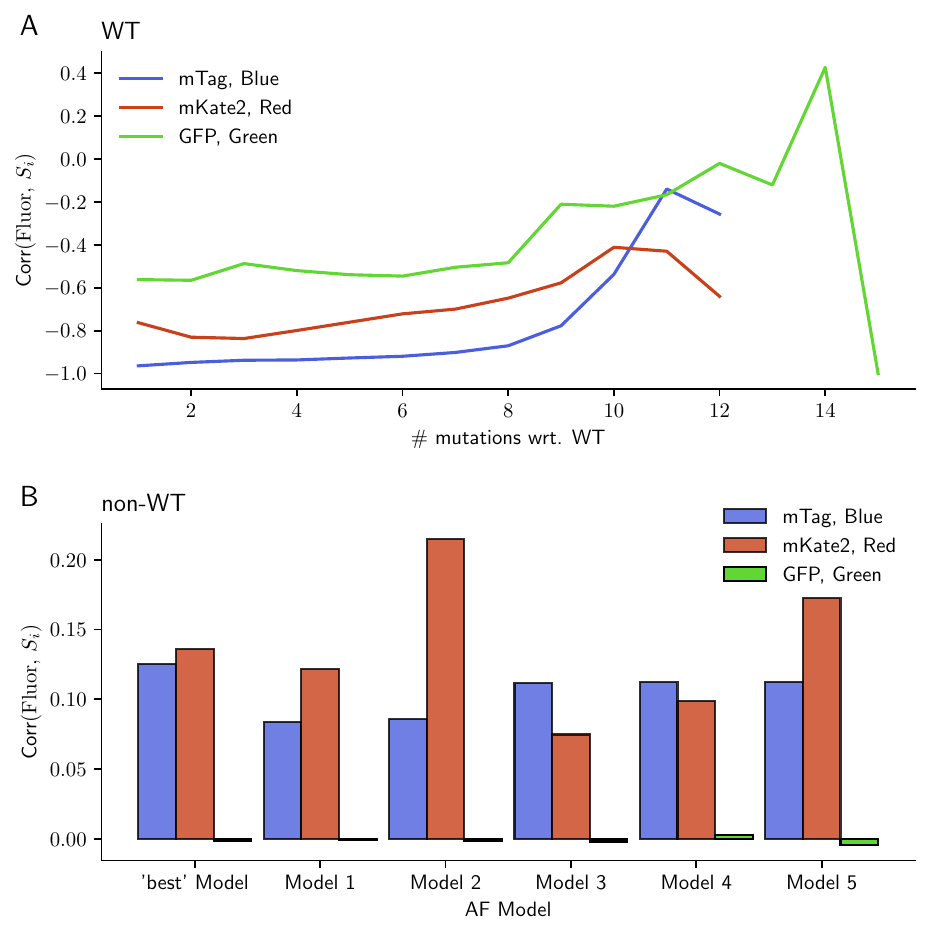}
  \caption{\label{fig:si6b}
  A: Correlation between phenotype and ES ($\gamma = \SI{13}{\angstrom}$) as a function of the number of mutations from the WT protein.
  B: Correlation (Pearson's $r$) between structural change and the magnitude of phenotypic change between two sequences, $i$ and $j$: $|\phi_i - \phi_j|$, where $\phi$ is the relevant phenotype.
  Correlations are shown for all AF models, for the effective strain at particular residues: mTag, $i=65$; mKate2, $i=218$; GFP, $i=58$.
  For each data set, correlations are obtained by comparing \num{10000} pairs of sequences chosen randomly with replacement.
  }
  \end{figure}

  \begin{figure}[t!]  \centering
  \includegraphics[width=0.99\linewidth,trim={0.5in 0.5in 0.5in 0.5in},clip]{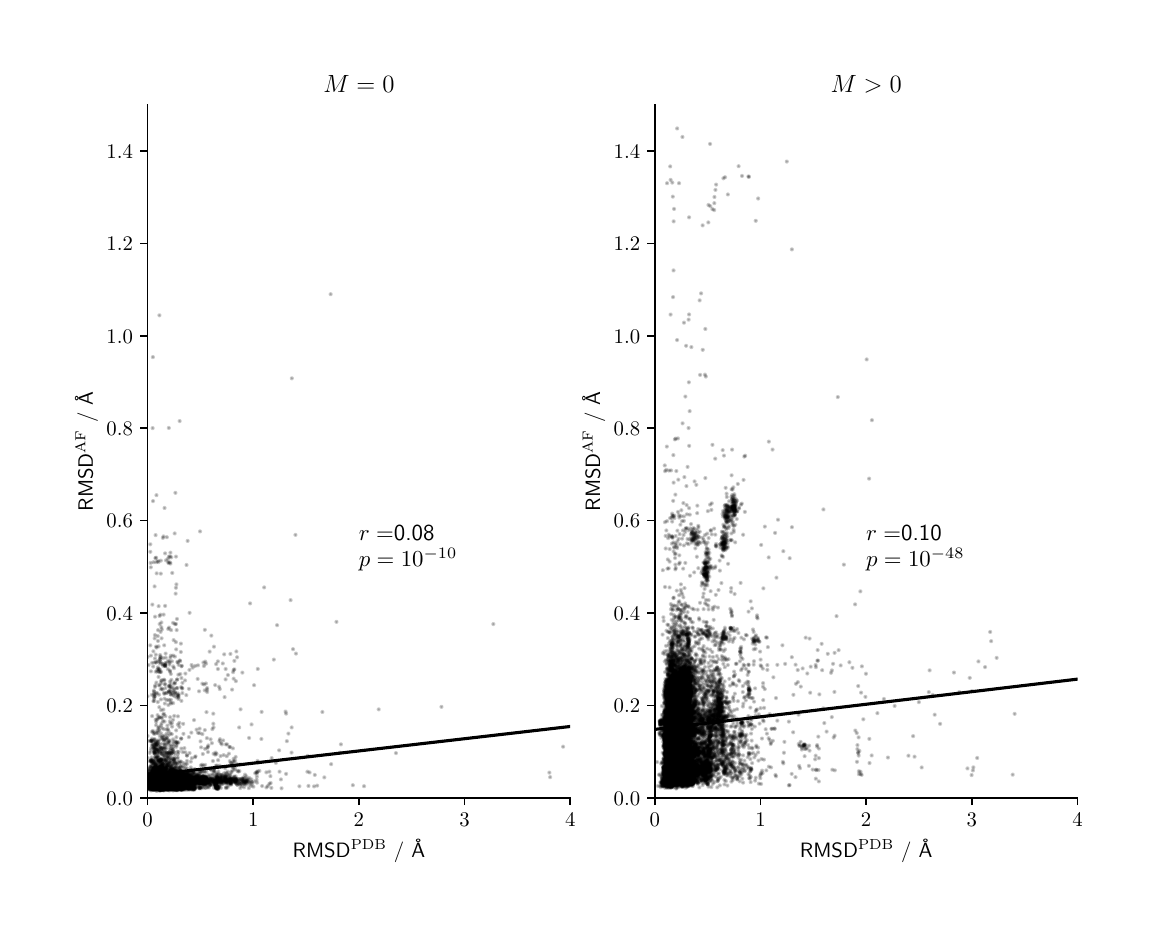}
  \caption{\label{fig:si9}
  \textbf{Correlations between empirical and predicted RMSD.}
  RMSD calculated using PDB structures, versus RMSD calculated using AF-predicted
  structures, for the full set of structure pairs from the PDB;
  non-mutated pairs (left) and mutated pairs (right) are shown separately.
  Pearson's correlation coefficient and linear fit are shown.
  }
  \end{figure}

  \subsection{Dependence on number of mutations}
  We consider the possibility that since deformation likely increases with
  the number of mutations, the phenotype-deformation correlations may be 
  simply due to an underlying correlation between phenotype and the number of
  mutations from the WT. To examine this, we measured correlations between phenotype and ES, controlling for the number of mutations (\fref{fig:si6b}A).
  We find that the correlations are independent of the number of mutations in the region $M\leq 8$. Even further away from the WT than $M=8$ we see a drop in correlations. Thus, the phenotype-ES correlations are not due to the number of mutations, but they are only strong for comparing sequences close to the WT.
  
  \subsection{Comparing mutants with mutants}
  We suspect that the high deformation-phenotype correlations we found are
  only possible when comparing WT with mutants. Our reasoning is that
  deformation should, by itself, not be sufficiently informative to predict
  changes in complex phenotypes. However, if one assumes that WT proteins
  are locally optimal, then any changes to the structure ought to have a
  deleterious effect on phenotype. To examine this, we calculate correlations
  between deformation and absolute differences in phentoype for \num{10000} randomly chosen pairs of mutants (\fref{fig:si6b}B). We find that the correlations are considerably lower than when comparing WT to mutants, in support of our conjecture.

  \begin{figure}[htb!]  \centering
  \includegraphics[width=0.99\linewidth]{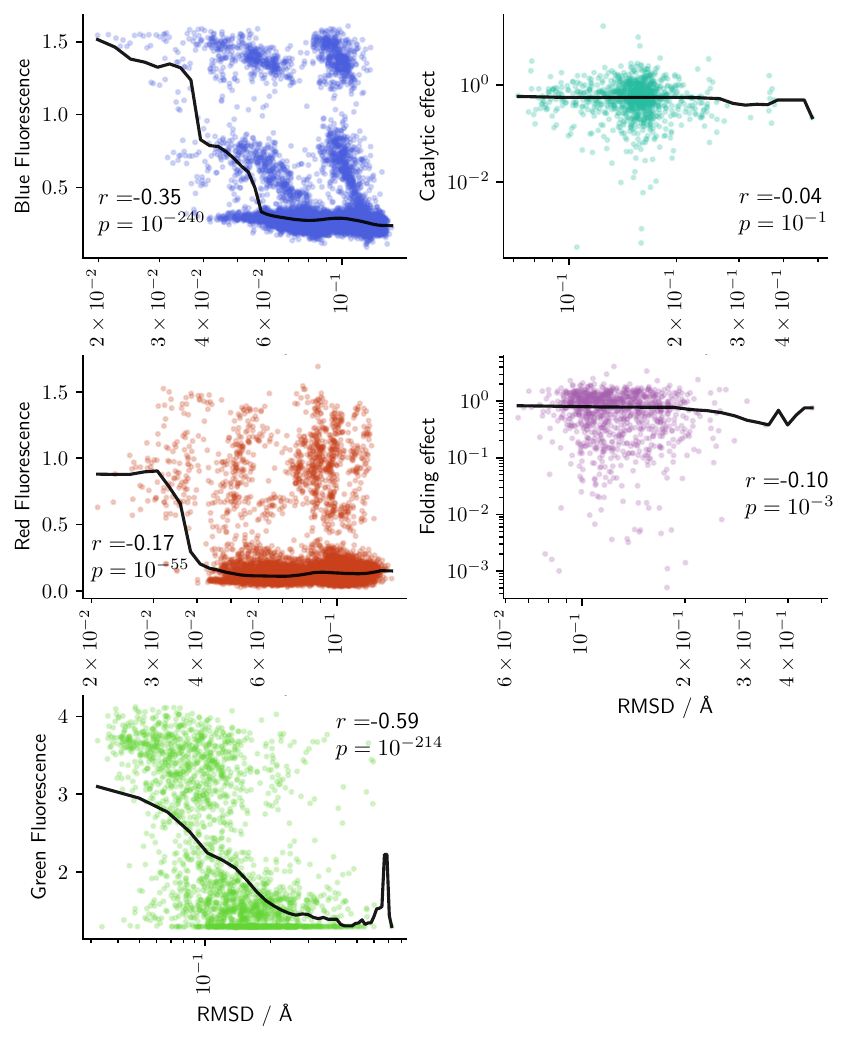}
  \caption{\label{fig:si10}
  \textbf{Correlations between RMSD and phenotype measurements.}
  Phenotype as a function of RMSD between WT and mutant.
  Red line indicates the median, calculated using a sliding window.
  Pearson's $r$ and corresponding $p$-values are shown on the graph.
  }
  \end{figure}

  \section{RMSD}

  At the outset, we expected that since mutation effects are assumed to 
  be typically localized, it would be prudent to use a local measure
  of deformation. Here we show that using RMSD, which depends on global alignment, is indeed insufficient
  to measure the effect of a mutation. We calculate the correlation between
  RMSD values obtained using PDB vs AF-predicted structures, for all
  structure pairs in our PDB dataset. \fref{fig:si9} shows that the correlation is quite
  weak, and does not differ for mutated vs non-mutated pairs.
  We also show correlations for the best-performing AF models of
  RMSD between WT and mutant structures, and the corresponding phenotype (\fref{fig:si10}).
  By its nature, RMSD is a global measure that obscures localized deformation effects. Thus, as expected, the correlations for RMSD are much lower than what
  is observed for local measures of deformation (\fref{fig:si4});
  one exception is GFP, where RMSD correlates almost as well with
  green fluorescence ($r=-0.59$). 


  \begin{figure}[t!]  \centering
  \includegraphics[width=0.99\linewidth]{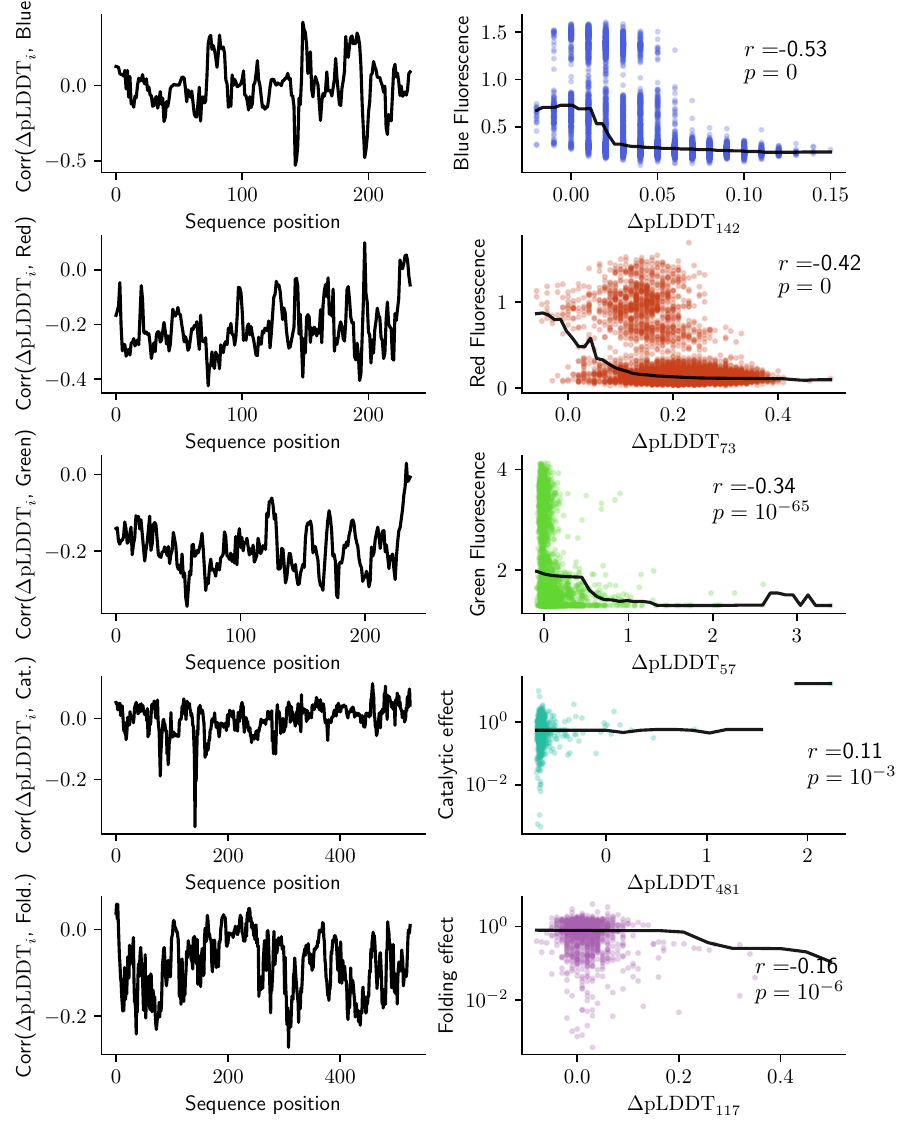}
  \caption{\label{fig:si11}
  \textbf{Correlations between $\Delta$pLDDT and phenotype measurements.}
  (Left) Correlations between $\Delta$pLDDT at residue $i$ and phenotype for \num{5} sets of
  phenotype measurements (top to bottom: mTag, blue fluorescence; mTag, red fluorescence; GFP, green fluorescence;
  PafA, catalytic effect; PafA, folding effect) as a function of sequence position $i$.
  In each case, results are shown for the AF model with the highest correlation.
  (Right) Phenotype as a function of $\Delta$pLDDT for the residue with the highest
  correlation. Red line indicates the median, calculated using a sliding window.
  Pearson's $r$ and corresponding $p$ values are shown on the graph; zero values 
  are shown for $p$ when $p$ is lower than is possible for 64-bit floating point precision.
  }
  \end{figure}

  \begin{figure}[t!]  \centering
  \includegraphics[width=0.99\linewidth,trim={0.5in 0in 0.5in 0.0in},clip]{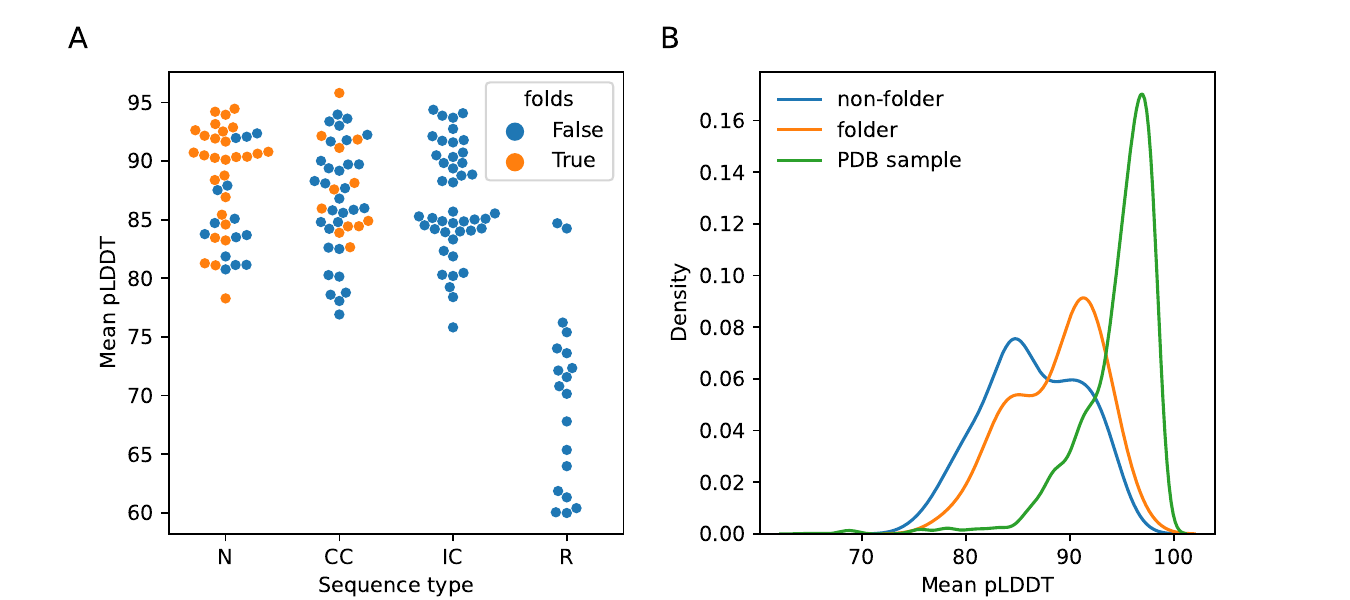}
  \caption{\label{fig:si13b}
  A: Mean pLDDT for 147 protein sequences based on the WW domain, some of which
  are found to fold \textit{in vitro}, and some of which do not fold.
  Proteins are grouped into: natural sequences (N); generated sequences
  that matches the covariance of the natural sequences (CC); generated
  sequences that match the statistics of the natural sequences at each position (IC);
  sequences generated by shuffling the natural sequences (R). 
  B: We compare mean pLDDT of the folding and non-folding WW domain sequences (excluding
  the shuffled sequences, as these share little sequence identity with the natural sequences),
  with a sample of mean pLDDT values for proteins in the PDB (same sampling procedure as in the main text).
  In this case, mean pLDDT appears to be of little use in differentiating folding and non-folding
  sequences.
  }
  \end{figure}


  \section{pLDDT}\label{sec:plddt}

  Since AlphaFold has been released, much attention has been given to pLDDT,
  AlphaFold's predicted (using a neural network) confidence score. pLDDT is, effectively, AlphaFold's attempt
  to predict the LDDT score between the predicted structure and the ground truth (PDB).
  Recent studies have shown that pLDDT is a good predictor of disorder~\cite{piops22},
  and that it is correlated with measures of flexibility:
  root-mean-square deviation measured in molecular dynamics simulations~\cite{guosr22};
  S2 from NMR~\cite{maps23}. This makes sense since flexible regions will inevitably
  lead to lower-than-average LDDT scores across repeat measurements or predictions.
  Although there is surely more to uncover, the evidence currently points to pLDDT as being a
  measure of confidence in the prediction of a residue's position, and
  a proxy measure for local flexibility. In this work, for example,
  we exclude residues from deformation calculations if they have pLDDT $< 70$.

  \subsection{Correlations between changes in pLDDT and phenotype}

  It is plausible that changes in pLDDT may correlate with phenotype, as large changes in
  pLDDT not only indicates changes in flexibility, but may also be an indicator of large deformation.
  We therefore look at AF-predicted changes in pLDDT for each residue in mutants compared to the wild type.
  We calculate the correlation between $\dP_i = \textrm{pLDDT(WT)} - \textrm{pLDDT(Mutant)}$
  and phenotype for all five sets of phenotype measurements (\fref{fig:si11}).
  We find statistically significant correlations between $\dP_i$ and phenotype, but in each case correlations are 
  about half of what can be achieved using other deformation metrics.

  Despite using similar data to \cite{pakpo23}, it might seem that we find higher correlations between pLDDT and fluorescence, but there are two important differences: In \cite{pakpo23} they study \num{447} single mutants, while we study \num{2312} mutants, \num{200} of which are single mutants. We calculate correlations between $\dP_i$ and fluorescence, while \cite{pakpo23} calculate the difference in pLDDT at the mutated site $\dP_m$ ($r = 0.17$), and the difference in mean pLDDT, $\Delta \langle \textrm{PLDDT} \rangle$ ($r = 0.16$). For a more direct comparison with this study, we calculated correlations between the same quantities on the \num{200} single mutants that we examined. Like \cite{pakpo23}, we found that these quantities are not very predictive of fluorescence ($\dP_m$, $r=0.03$; $\Delta \langle \textrm{PLDDT} \rangle$, $r = 0.06$).

  \subsection{Is pLDDT a good predictor of protein folding?}
  We are quite confident that pLDDT is a good measure of local flexibility, 
  and that values under 70 are strong indicators of disorder. However,
  it is less clear whether average (across all residues) pLDDT values
  are useful for predicting whether a protein will fold or not, 
  especially when the average is above 70. To test this, we looked at the mean
  pLDDT for a set of WW-domain-like proteins from \cite{socna05} (\fref{fig:si13b}).
  We find that proteins that were created from randomly shuffled sequences
  produced noticeably lower mean pLDDT values than the other, less random
  sequences. Proteins that had mutations that only took into consideration
  the positional statistics (IC; \ie, the likelihood of an amino acid at
  a certain position in the sequence, as determined from a multiple sequence
  alignment) were not found to fold \textit{in vitro}, yet they have
  comparable values of mean pLDDT to the WT sequences that were tested (natural sequences, N).
  This suggests that mean pLDDT alone is not a useful predictor of whether a protein will fold.

  \section{AlphaFold Models}\label{sec:model}

\begin{figure}[t!]  \centering
  \includegraphics[width=0.99\linewidth]{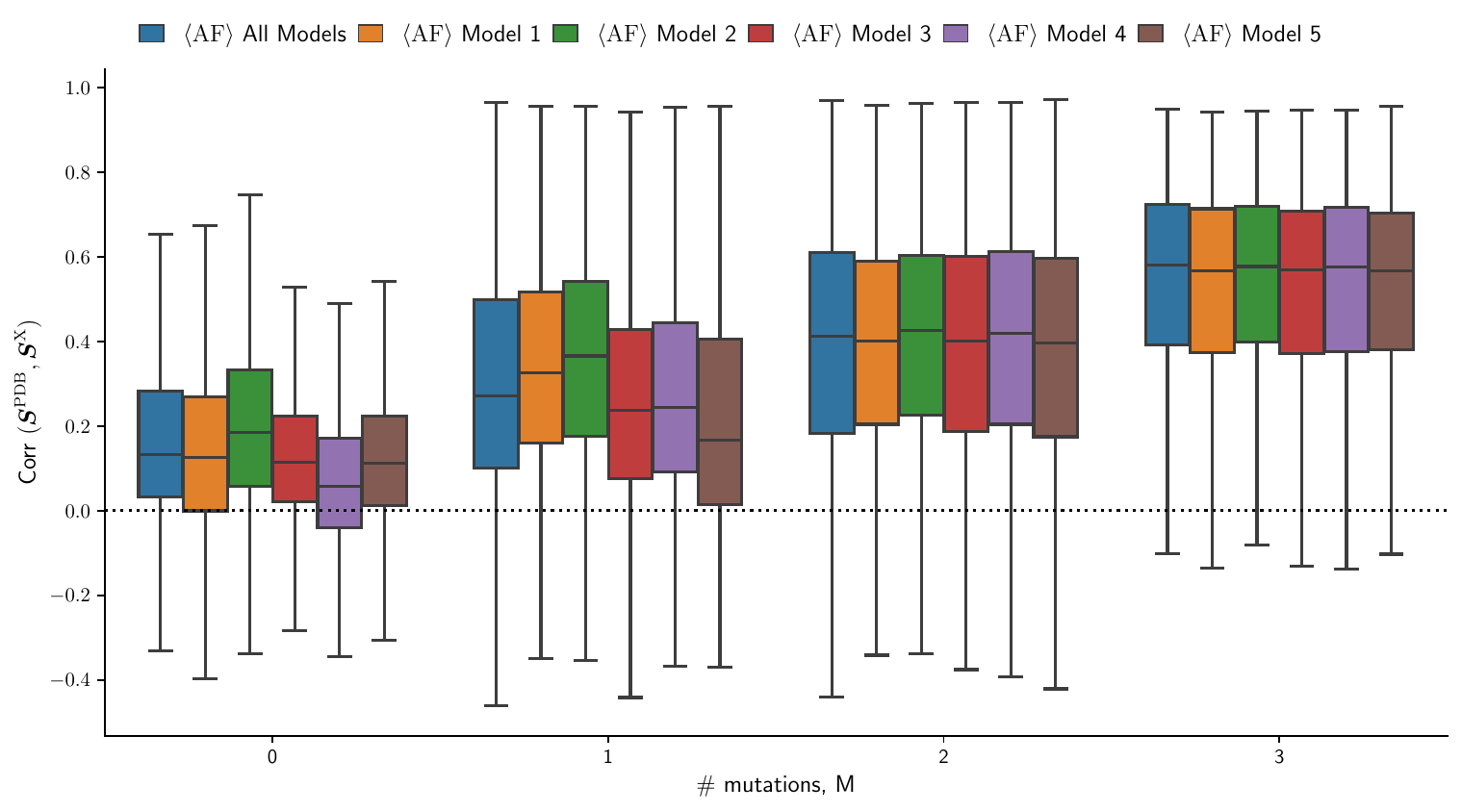}
  \caption{\label{fig:si12a}
  \textbf{Effect of AF model.}
  PDB - $\AFave$ correlation as a function of $M$ and AF model type.
  }
  \end{figure}

  Since AF was released with \num{5} sets of trained model weights, there is a choice of which structure to use. We find that no one model produces significantly better correlations than the other models (\fref{fig:si12a}).
  We find significant differences between the predictions of different models on correlations with phenotype (\fref{fig:si4}),
  but the scale of these differences are reduced by calculating average structures ($\AFave$).
  We found one case (PafA) where models \num{1} and \num{2} produced poor predictions (\fref{fig:si4}).
  In this case, most of the sequence variants were predicted to have very similar structures.
  We attribute this odd behavior to the use of structural templates, which are used by models \num{1} and \num{2}
  by default. We find much better results for these two models when using the ColabFold implementation
  (\fref{fig:si7}, \fref{fig:si8}), where we were able to run AF without using structural templates.
  This leads leads us to recommend not using structural templates in AF predictions.
  In general, the repeat predictions of structures are more similar when produced by the same model
  (and conversely, different models produce different structures to each other);
  for this reason we also advise against averaging over structures from different models
  unless there are multiple repeat predictions from each model (in our case, we used \num{6} repeat
  structures to create average structures).

  \section{Scale of mutation effects}

  \begin{figure}[t!]  \centering
  \includegraphics[width=0.99\linewidth]{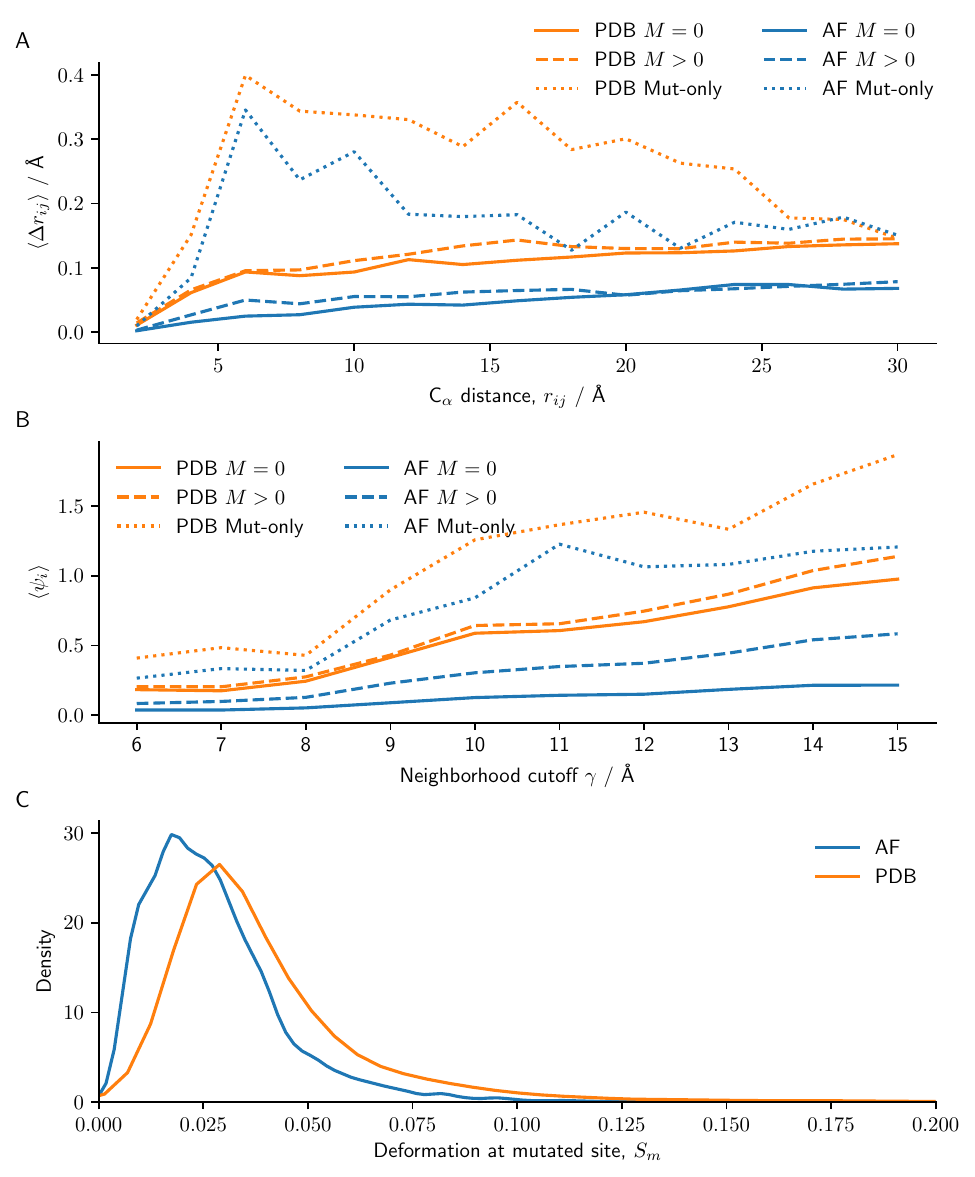}
  \caption{\label{fig:si12}
  \textbf{Changes in backbone distances, neighborhoods and deformation due to mutation.}
  A: Average change in $\CA$ distances between structures, as a function of $\CA$ distance.
  B: Average neighbor mismatch, $\psi_i$, as a function of neighborhood cutoff $\gamma$.
  Separate lines are shown for PDB and AF structures, and for all pairs
  of residues in non-mutated pairs ($M=0$), all pairs of residues in mutated
  pairs ($M>0$), and only comparing residues with mutated residues (Mut-only).
  C: Distribution of mutation effects, $S_\mathrm{m}$, for PDB and AF-predicted pairs
  of structures.
  }
  \end{figure}

  \subsection{Deformation between PDB structures is typically small}
  To give an alternative view of the magnitude of deformation in the PDB, we examine
  changes to distances between residues.
  We calculate the distance between all $\CA$ positions, $r_{ij}$,
  and then get the absolute difference between these distances in a reference structure
  and a target structure, $\Delta r_{ij} = r_{ij} - r_{ij}'$.
  In \fref{fig:si12}A, we show the mean absolute difference between $\CA$ distances, as
  a function of $\CA$ distance; we calculate this for all residues in all protein pairs
  with $M=0$ and $M>0$, and also for $\CA$ distances with respect to mutated residues.
  On average, backbone $\CA$ positions move by less than \SI{0.2}{\AA}, both for
  PDB and AF-predicted structures. However, when looking at backbone distances only from
  mutated sites, we see average mutations of up to \SI{0.4}{\AA} within \SI{1}{\nm}
  from mutated sites. It seems that a few mutations typically lead to rather 
  slight changes to bulk protein structure, yet can have significant effects locally.
  
  When we calculate deformation, we include in our calculation residues that are
  neighbors in both structures. One might suspect that this can lead to problems if
  deformation leads to big changes in neighborhoods. To test this, for each residue
  we calculate the symmetric neighbor difference, $\psi_i$, as the number of residues that are not shared between neighbor sets in a reference, $N_i$
  and target structure, $N'_i$,
  \be
  \psi_i = \left| N_i \ominus N'_i \right|~.
  \ee
  We calculate the average neighbor difference across all pairs and residues, $\langle \psi_i \rangle$, as a function of neighborhood
  cutoff $\gamma$. We show in \fref{fig:si12}B that
  neighbors typically differ by less than one residue, even when focusing on mutated
  sites.

  To help put mutation effects in context, we plot the distribution of 
  mutation effects (ES at the mutated site, $S_\mathrm{m}$)
  for both PDB and AF-predicted structures (\fref{fig:si12}C). This provides a handy
  reference for understanding when an effect is relatively large or small.

  \begin{figure}[t!]  \centering
  \includegraphics[width=0.99\linewidth]{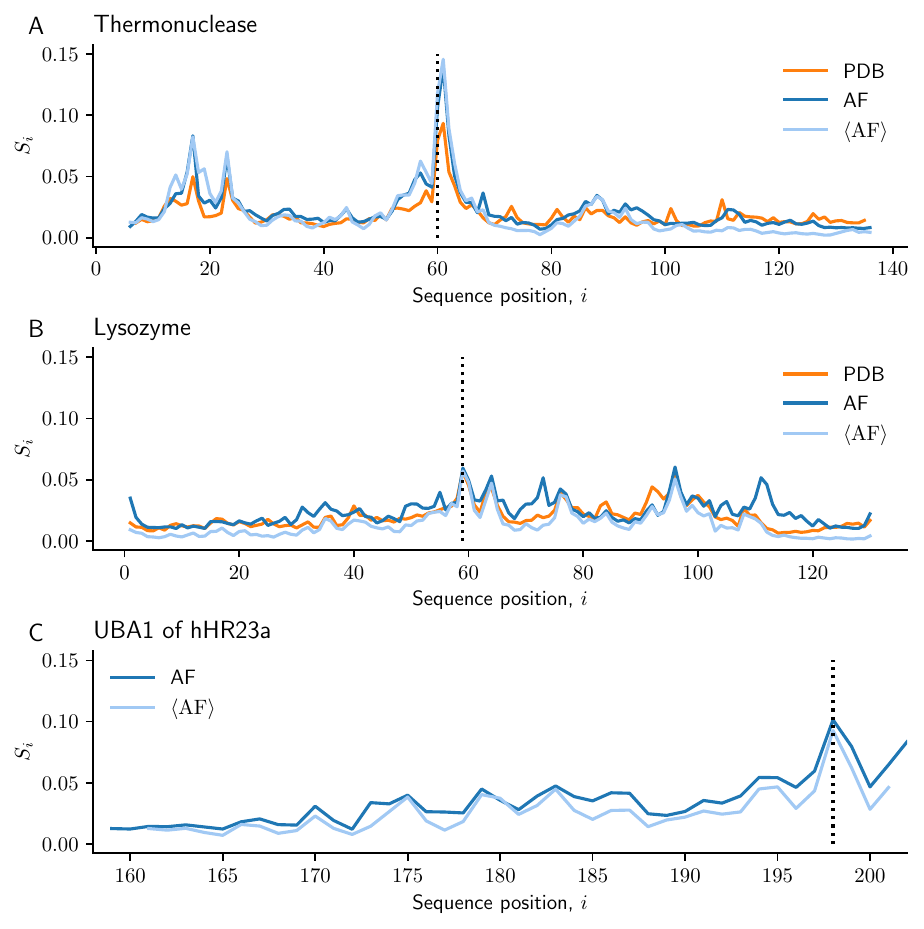}
  \caption{\label{fig:si13}
  \textbf{Examples of large deformation.}
  Deformation per residue for three examples: Thermonuclease from \textit{staphylococcus aureus} (A),
  human Lysozyme (B), and the UBA1 domain of human hHR23a (C). Deformation is shown for
  AF, $\AFave$, and PDB structures where possible. The location of mutations is indicated
  by dotted lines.
  }
  \end{figure}

  \subsection{Example of large deformation}
\begin{figure*}[t!]  \centering
  \includegraphics[width=0.99\linewidth]{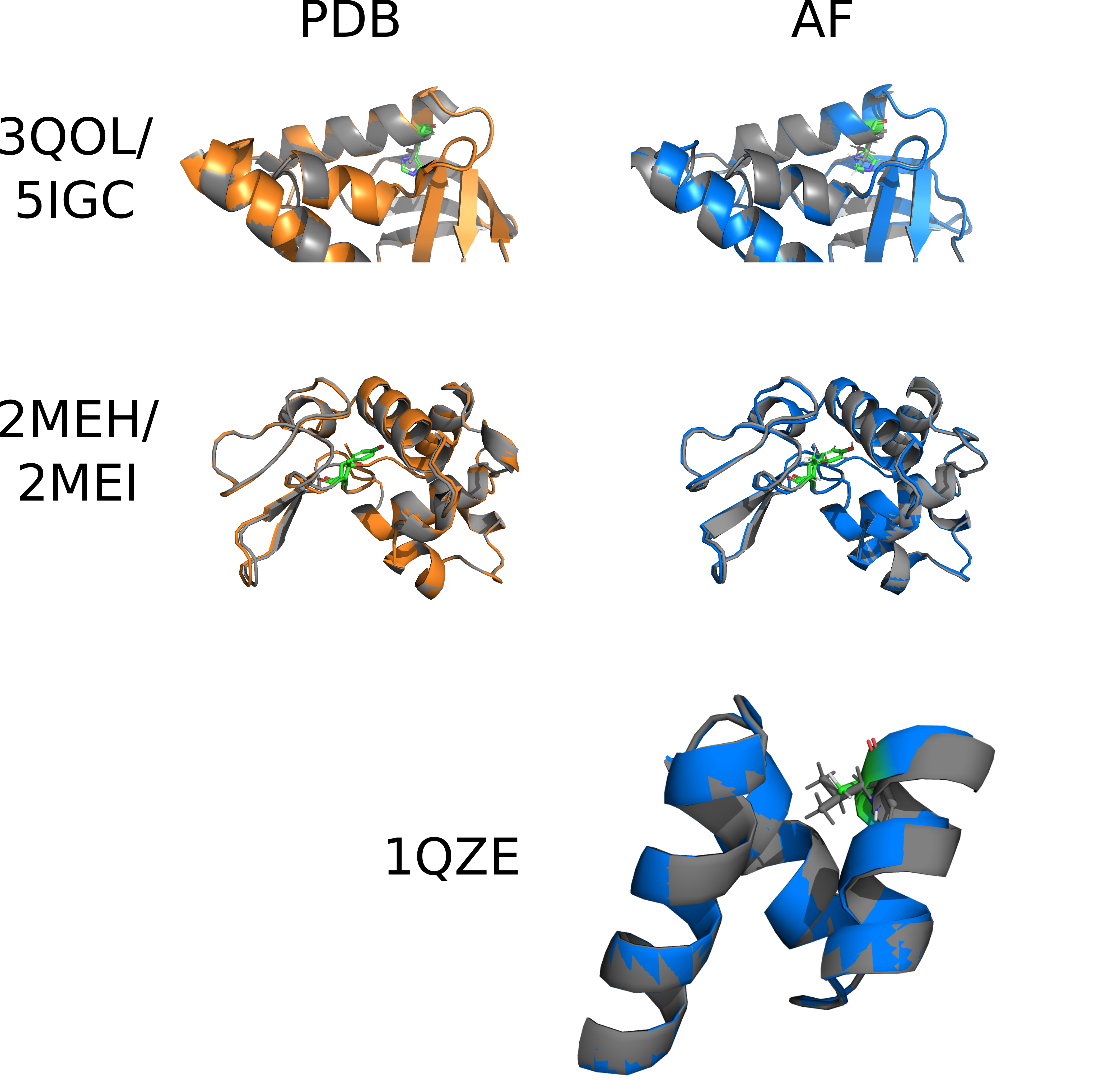}
  \caption{\label{fig:si14}
  \textbf{Examples of large deformation.}
  Comparisons of PDB and AF-predicted structures that differ by one amino acid:
  Thermonuclease from \textit{staphylococcus aureus} (V60H, \texttt{3QOL\_A} and \texttt{5IGC\_A}; a close-up of the region affected by the mutation is shown),
  human Lysozyme (T59Y, \texttt{2MEH\_A} and \texttt{2MEI\_A}), and the UBA1 deomain of human hHR23a (L198A).
  Mutated structures are aligned and shown overlaid as cartoons; atomic positions are shown for the mutated residues.
  PDB structures are only shown for examples where both proteins are available.
  }
  \end{figure*}

  Typically, mutation effects in the PDB are quite small, such that they often
  have only subtle effects even on local structure. Perhaps the best
  way to illustrate this is to show examples of some of the largest deformations
  measured. We chose two examples of proteins that differ by one amino acid,
  which exhibit some of the highest deformations found in
  the PDB: (i) thermonuclease (V60H, comparing \texttt{3QOL\_A} and \texttt{5IGC\_A})) and (ii) lysozyme (T59Y, comparing \texttt{2MEH\_A} and \texttt{2MEI\_A}). In each case, while deformation is higher (\fref{fig:si13})
  then what is typical (\fref{fig:si12}C), the structural differences are not so severe (\fref{fig:si14}).
  In thermonuclease, the replacement of a valine with histidine results in a slight kink in the
  $\alpha$-helix, due to the greater size of the histidine which interferes with how the
  amino acids pack. In lysozyme, we see a similar effect but this time in a more central location
  within the protein; replacing threonine with tyrosine results in the neighboring $\alpha-$helix
  and $\beta-$sheets being pushed apart. Both of these effects are recapitulated
  in AF predictions (\fref{fig:si14}). We note that mutation effects of this magnitude
  are rare in the PDB, yet they are still difficult to evaluate visually. Nor do we find pLDDT
  to be a reliable indicator of structure change: thermonuclease, $\dP_m=-2.5$; lysozyme, $\dP_m=0.04$.
  Instead, \emph{measuring deformation allows us to quantify the effect of a mutation with much higher precision,
  in a way that can be directly compared with experimental results}.
  
  \subsection{UBA1 of hHR23a}
  A recent paper studied the UBA1 domain of hHR23a protein (amongst other examples), and suggested
  that differences in packing, and differences in pLDDT
  constitute evidence that AF cannot predict the effect of missense mutations~\cite{buens22}.
  We here show that the effect of the mutation (L198A) is actually quite large compared
  to average effects (\fref{fig:si13}C). The deformation at the mutated site is $S_\mathrm{m}=0.093$,
  which is higher than \SI{97}{\%} of all deformation values measured at mutated sites in PDB structures.
  The mutation in question leads to higher degradation by proteases, which shows
  that this mutation destablizes the structure. We consider that such large deformations
  are perhaps likely to destabilize the structure, given how rare they are in the PDB.
  The PDB is constructed with a sampling bias towards proteins that can fold, so the
  fact that there are so few mutants with this level of deformation might indicate
  that high deformation is a good predictor of destabilization. This is an interesting 
  area for future study.

  \begin{figure*}[t!]  \centering
  \includegraphics[width=0.80\linewidth]{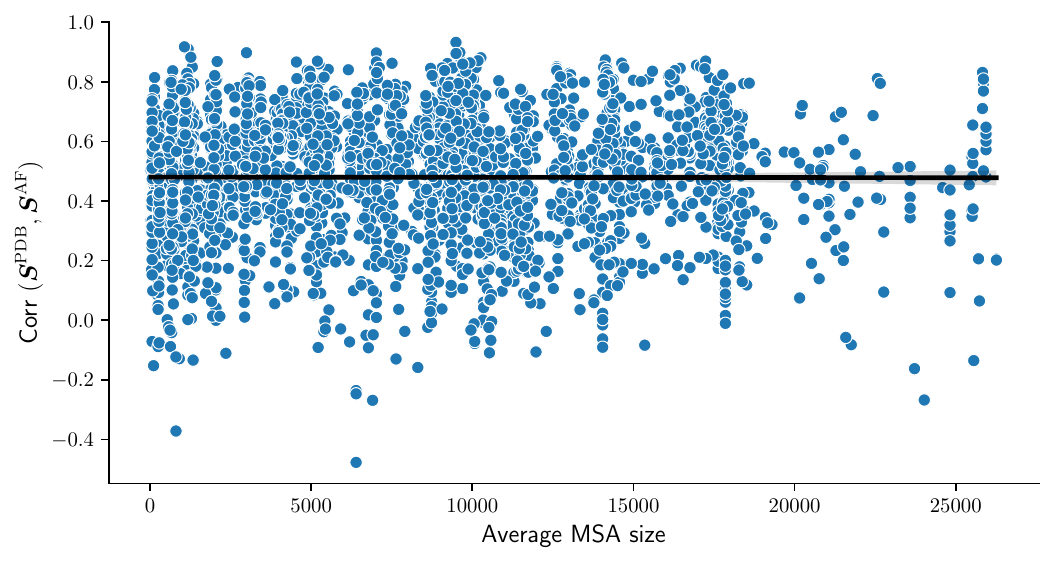}
  \caption{\label{fig:si21}
  \textbf{Effect of MSA size on PDB-AF correlation.}
  Average MSA size is the average across the two proteins in each pair of mutants. For protein pairs with no mutation, it is just the size of the MSA. Results are shown for a non-redundant sample.
  }
  \end{figure*}

 \clearpage

\bibliography{master}